# Electromagnetic-Power-based Characteristic Mode Theory for Perfect Electric Conductors

Renzun Lian

*Abstract*—In this paper, a new way to normalize various electromagnetic powers of PEC systems and a new way to decompose power quadratic matrix are introduced, and then an ElectroMagnetic-Power-based Characteristic Mode Theory (CMT) for PEC systems (PEC-EMP-CMT) is built.

The PEC-EMP-CMT is not only valid for the PEC systems which are placed in vacuum, but also suitable for the PEC systems which are surrounded by any electromagnetic environment. Based on the new normalization way and the new decomposition way, the complex characteristic currents and non-radiative Characteristic Modes (CMs) can be constructed; some traditional concepts, such as the system input impedance and modal input impedance etc., are redefined based on the language of field instead of the language of circuit; the traditional characteristic quantity, Modal Significance (MS), is generalized; a series of new power-based CM sets, such as the Radiated power CM (RaCM) set and Output power CM (OutCM) set etc., are introduced.

It is proven in this paper that various power-based CM sets of a certain objective PEC structure are independent of the external environment and excitation; the OutCM set has the same physical essence as the CM set derived from the traditional Surface EFIE-based CMT for PEC systems (PEC-SEFIE-CMT), but the former can be viewed as the generalization of the latter; the non-radiative space constituted by all non-radiative modes is identical to the interior resonance space constituted by all interior resonant modes of closed PEC structures, and the non-radiative CMs constitute a basis of the space.

Based on above these, the normal Eigen-Mode Theory (EMT) for closed PEC structures and the PEC-SEFIE-CMT are classified into the PEC-EMP-CMT framework, and then the ideas of PEC-SEFIE-CMT are liberated from its traditional framework, MoM. In addition, a variational formulation for the external scattering problem of PEC structures is provided in this paper, based on the conservation law of energy.

## I. Introduction

THE Theory of Characteristic Mode (TCM), or equivalently called as Characteristic Mode Theory (CMT), was firstly introduced by Robert J. Garbacz [1], and subsequently refined by Roger F. Harrington and Joseph R. Mautz under the Integral Equation-based MoM (IE-MoM) framework. In 1971, Harrington and Mautz built their CMT for PEC systems based on the Surface EFIE-based MoM (PEC-SEFIE-CMT) [2]-[3]. Afterwards, some variants for the PEC-SEFIE-CMT were introduced under the IE-MoM framework, such as the Volume Integral Equation-based CMT for Material bodies (Mat-VIE-CMT) [4], the Surface Integral Equation-based CMT for Material bodies (Mat-SIE-CMT) [5], the Surface MFIE-based CMT for PEC systems (PEC-SMFIE-CMT) [6], and the Surface CFIE-based CMT for PEC systems (PEC-SCFIE-CMT) [7]. Recently, the PEC-SEFIE-CMT is re-derived from complex Poynting's theorem in [8], and the power characteristics of the Characteristic Modes (CMs) constructed by Mat-SIE-CMT is researched in [9], such that the physical pictures of the PEC-SEFIE-CMT and Mat-SIE-CMT become clearer.

In fact, the CMT is a new modal theory which is essentially different from the traditional Eigen-Mode Theory (EMT) [10]-[12] in mathematical physics. The EMT is a source-free modal theory or called as passive modal theory, but the CMT is an active modal theory. In mathematical physics, the EMT is divided into two categories, the normal EMT (such as the passive modal problems of closed structures [11]) and the singular EMT (such as the passive modal problems of open structures [12]). Generally speaking, the active modal theory is more suitable for analyzing the open electromagnetic systems, because the Sommerfeld's radiation condition [13] at infinity signifies a continuous energy output from the open system, and then a non-zero external excitation is necessary for the system to sustain a stationary oscillation of field. Because of these above, the CM-based analysis and design methodologies for various open electromagnetic systems, especially for radiation systems, have got more and more attentions in electromagnetic engineering society, and a comprehensive review for the CMT and its applications in antenna community can be found in [8].

In this paper, a new way to normalize various electromagnetic powers of PEC systems and a new way to decompose power quadratic matrix are introduced, and then an ElectroMagnetic-Power-based CMT for PEC systems (PEC-EMP-CMT) is established. The PEC-EMP-CMT is not only valid for the PEC systems which are placed in vacuum, but also suitable for the ones which are surrounded by any electromagnetic environment.

Based on the new normalization way and the new decomposition way, the complex characteristic currents and non-radiative CMs can be constructed; some traditional concepts, such as the system input impedance and modal input impedance etc., are redefined based on the language of field instead of the language of circuit; the traditional characteristic

R. Z. Lian is with the School of Electronic Engineering, University of Electronic Science and Technology of China, Chengdu 611731, China. (e-mail: rzlian@vip.163.com).

quantity, Modal Significance (MS) [8], is generalized; a series of power-based CM sets, such as the Radiated power CM (RaCM) set and Output power CM (OutCM) set etc., are introduced.

The various power-based CM sets of a certain objective PEC structure are independent of the external environment and excitation. Among these CM sets, the OutCM set has the same physical essence as the CM set derived from PEC-SEFIE-CMT, but the former can be viewed as the generalization of the latter, and then the ideas of PEC-SEFIE-CMT are liberated from its traditional framework, MoM.

It is proven in this paper that all non-radiative modes constitute a linear space called as non-radiative space, and the space is identical to the interior resonance space constituted by all interior resonant modes of closed PEC structures, and the non-radiative CMs derived in PEC-EMP-CMT constitute a basis of the space. Based on above these, the normal EMT for closed PEC structures [11] is classified into the PEC-EMP-CMT framework.

In addition, a variational formulation for the external scattering problem of PEC structures is provided in this paper, based on the conservation law of energy [14].

This paper is organized as follows. Sections II-VII give the principles and formulations related to PEC-EMP-CMT, and some necessary discussions are provided in Sec. VIII. Section IX concludes this paper. In what follows, the $e^{j\omega t}$ convention is used throughout.

## II. VARIOUS ELECTROMAGNETIC POWERS AND THEIR NORMALIZATIONS

In this paper, the power characteristics of a PEC system which is placed in an arbitrary time-harmonic electromagnetic environment is studied. When an external excitation is impressed, there exist three kinds of fields in whole space $\mathbb{R}^3$, that are the $\bar{F}^{im}$ generated by impressed excitation, the $\bar{F}^{en}$ generated by external environment, and the $\bar{F}^{sys}$ generated by PEC system, as illustrated in Fig. 1, here $F = E, H$. The term "time-harmonic electromagnetic environment" means that the $\bar{F}^{en}$ is a time-harmonic electromagnetic field operating at the same frequency as $\bar{F}^{im}$.

Based on the superposition principle [14], the $\bar{F}^{sys}$ is considered as the scattering field due to the incident field $\bar{F}^{inc} = \bar{F}^{im} + \bar{F}^{en}$, because the PEC system is regarded as a whole object in this paper. The $\bar{F}^{sys}$ is equivalently written as $\bar{F}^{sca} = \bar{F}(\bar{J}^{sca})$, and the PEC system is simply called as scatterer. The $\bar{J}^{sca}$ is the surface scattering current on the scatterer, and the symbol $\bar{F}(\bar{J}^{sca})$ represents the field generated by $\bar{J}^{sca}$ in vacuum, and the mathematical expression for this operator can be found in [13].

### A. Various electromagnetic powers.

The power done by $\bar{F}^{inc}$ on $\bar{J}^{sca}$ is just the input power for scatterer from both impressed excitation and external environment, and its mathematical expression is as follows

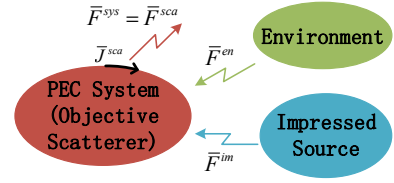

Fig. 1. Various fields generated by various sources.

$$\begin{aligned} P^{inp}(\bar{J}^{sca}) &= \frac{1}{2}\langle \bar{J}^{sca}, \bar{E}^{inc}\rangle_{\partial V} \\ &= -\frac{1}{2}\langle \bar{J}^{sca}, \bar{E}(\bar{J}^{sca})\rangle_{\partial V} \quad (1) \\ &= P_{rad}^{sca}(\bar{J}^{sca}) + j2\omega[W_m^{sca}(\bar{J}^{sca}) - W_e^{sca}(\bar{J}^{sca})] \end{aligned}$$

In (1), the coefficient "1/2" is due to time average; the $\omega = 2\pi f$ is angle frequency, and the $f$ is frequency; the inner product is defined as $\langle \bar{f}, \bar{g}\rangle_\Omega \triangleq \int_\Omega \bar{f}^* \cdot \bar{g}\, d\Omega$, and the symbol "·" is the dot multiplication of field vectors, and the superscript "∗" represents complex conjugate; the $\partial V$ is the whole boundary of scatterer $V$. The second and third equalities in (1) respectively originate from the SEFIE [13] and complex Poynting's theorem [15]. In (1), the $P_{rad}^{sca}$, $W_m^{sca}$, and $W_e^{sca}$ are respectively the radiated power carried by $\bar{F}^{sca}$ and the total magnetically and electrically stored energies of $\bar{F}^{sca}$, and [15]

$$\begin{aligned} P_{rad}^{sca}(\bar{J}^{sca}) &= \frac{1}{2}\oiint_{S_\infty}[\bar{E}^{sca} \times (\bar{H}^{sca})^*] \cdot d\bar{S} \\ &= \frac{1}{2\eta_0}\langle \bar{E}^{sca}, \bar{E}^{sca}\rangle_{S_\infty} \quad (2.1) \\ &= \frac{\eta_0}{2}\langle \bar{H}^{sca}, \bar{H}^{sca}\rangle_{S_\infty} \end{aligned}$$

$$W_m^{sca}(\bar{J}^{sca}) = \frac{1}{4}\langle \bar{H}^{sca}, \mu_0 \bar{H}^{sca}\rangle_{\mathbb{R}^3} \quad (2.2)$$

$$W_e^{sca}(\bar{J}^{sca}) = \frac{1}{4}\langle \varepsilon_0 \bar{E}^{sca}, \bar{E}^{sca}\rangle_{\mathbb{R}^3} \quad (2.3)$$

here the relations $\bar{H}^{sca} = (1/\eta_0)\hat{n} \times \bar{E}^{sca}$ and $\bar{E}^{sca} = \eta_0 \bar{H}^{sca} \times \hat{n}$ on $S_\infty$ have been employed in the (2.1), and the $\eta_0 = \sqrt{\mu_0/\varepsilon_0}$ and $\hat{n}$ are respectively the wave impedance in vacuum and the outer normal of $S_\infty$; the $S_\infty$ is a closed spherical surface at infinity; the $\varepsilon_0$ and $\mu_0$ are respectively the permittivity and permeability in vacuum.

In fact, the system output power $P^{out}$ is as follows

$$\begin{aligned} P^{out}(\bar{J}^{sca}) &= P_{rad}^{sca}(\bar{J}^{sca}) + j2\omega[W_m^{sca}(\bar{J}^{sca}) - W_e^{sca}(\bar{J}^{sca})] \\ &= -\frac{1}{2}\langle \bar{J}^{sca}, \bar{E}(\bar{J}^{sca})\rangle_{\partial V} \end{aligned} \quad (3)$$

and the physical essence of the second equality in (1) is just the conservation law of energy [14], i.e.,

$$P^{out}(\bar{J}^{sca}) = P^{inp}(\bar{J}^{sca}) \quad (4)$$

The physical meanings of the real and imaginary parts of $P^{inp}$ are respectively the active and reactive powers due to the interaction between $\bar{F}^{inc}$ and scatterer. The physical meanings of the real and imaginary parts of $P^{out}$ are respectively the radiated power and reactively stored power generated by $\bar{J}^{sca}$. To simplify the symbolic system and to consider of the physical meanings of various powers mentioned above, the $\text{Re}\{P^{inp}\}$ and $\text{Im}\{P^{inp}\}$ are simply denoted as $P^{inp}_{act}$ and $P^{inp}_{react}$ respectively, and the $\text{Re}\{P^{out}\}$ and $\text{Im}\{P^{out}\}$ are simply denoted as $P^{out}_{rad}$ and $P^{out}_{sto}$ respectively, and the $P^{out}_{sto}$ is simply called as "stored power", so

$$P^{inp}_{act} = \text{Re}\{P^{inp}\} = \text{Re}\{P^{out}\} = P^{out}_{rad} = P^{sca}_{rad} \quad (5.1)$$

$$P^{inp}_{react} = \text{Im}\{P^{inp}\} = \text{Im}\{P^{out}\} = P^{out}_{sto} = 2\omega(W^{sca}_m - W^{sca}_e) \quad (5.2)$$

Obviously, there exist above equivalent relations between the symbolic systems $\{P^{inp}, P^{inp}_{act}, P^{inp}_{react}; input, active, reactive\}$ and $\{P^{out}, P^{out}_{rad}, P^{out}_{sto}; output, radiated, stored\}$, and the two symbolic systems have their own merits to depict the characteristics of scatterer from different aspects. The former symbolic system focuses on emphasizing the interaction between the $\bar{F}^{inc}$ and scatterer, whereas the latter symbolic system focuses on emphasizing the aspects of scatterer's ability to output various electromagnetic energies. In the following discussions, the latter symbolic system is utilized.

*B. The normalizations for various electromagnetic powers.*

The value of $P^{out}$ can be uniquely obtained from the $\bar{J}^{sca}$, and this is why the $P^{out}$ is expressed as the one-variable function of $\bar{J}^{sca}$ in (3). Because the $\bar{J}^{sca}(\bar{r})$ is a complex vector function of the position $\bar{r}$, here $\bar{r} \in \partial V$, so the $P^{out}$ depends on both the amplitude of $\bar{J}^{sca}$ and the distribution of $\bar{J}^{sca}$. The distribution of $\bar{J}^{sca}$ includes three aspects, that are the distribution of $|\bar{J}^{sca}|$, the distribution of the phase of $\bar{J}^{sca}$, and the distribution of the polarization of $\bar{J}^{sca}$ [8]. To eliminate the influence from the amplitude and to take account of the distribution and to consider of that $P^{out}(0) = 0$, the $P^{out}$ is normalized as follows

$$\tilde{P}^{out}(\bar{J}^{sca}) \triangleq \begin{cases} \dfrac{P^{out}(\bar{J}^{sca})}{(1/2)\langle \bar{J}^{sca}, \bar{J}^{sca}\rangle_{\partial V}} & , (\bar{J}^{sca} \neq 0) \\ 0 & , (\bar{J}^{sca} = 0) \end{cases} \quad (6.1)$$

$$= \tilde{P}^{out}_{rad}(\bar{J}^{sca}) + j\, \tilde{P}^{out}_{sto}(\bar{J}^{sca})$$

here

$$\tilde{P}^{out}_{rad}(\bar{J}^{sca}) = \begin{cases} \dfrac{P^{out}_{rad}(\bar{J}^{sca})}{(1/2)\langle \bar{J}^{sca}, \bar{J}^{sca}\rangle_{\partial V}} & , (\bar{J}^{sca} \neq 0) \\ 0 & , (\bar{J}^{sca} = 0) \end{cases} \quad (6.2)$$

$$\tilde{P}^{out}_{sto}(\bar{J}^{sca}) = \begin{cases} \dfrac{P^{out}_{sto}(\bar{J}^{sca})}{(1/2)\langle \bar{J}^{sca}, \bar{J}^{sca}\rangle_{\partial V}} & , (\bar{J}^{sca} \neq 0) \\ 0 & , (\bar{J}^{sca} = 0) \end{cases} \quad (6.3)$$

The $\tilde{P}^{out}$, $\tilde{P}^{out}_{rad}$, and $\tilde{P}^{out}_{sto}$ in (6) are respectively called as the normalized system output, radiated, and stored powers. In fact, their dimensions are Ohm, so they can also be called as system output impedance, radiated resistance, and stored reactance respectively, and then they can be equivalently denoted as $Z$, $R$, and $X$ as follows

$$Z(\bar{J}^{sca}) = \tilde{P}^{out}(\bar{J}^{sca}) \quad (7.1)$$

$$R(\bar{J}^{sca}) = \tilde{P}^{out}_{rad}(\bar{J}^{sca}) \quad (7.2)$$

$$X(\bar{J}^{sca}) = \tilde{P}^{out}_{sto}(\bar{J}^{sca}) \quad (7.3)$$

In addition, considering of the (4) and (5), the $Z$, $R$, and $X$ can also be called as system input impedance, resistance, and reactance respectively.

### III. THE MATRIX FORMS FOR VARIOUS POWERS

The $\bar{J}^{sca}$ can be expanded in terms of the basis function set $\{\bar{b}_\xi\}_{\xi=1}^{\Xi}$ as follows

$$\bar{J}^{sca}(\bar{a}) = \sum_{\xi=1}^{\Xi} a_\xi \bar{b}_\xi = \bar{B} \cdot \bar{a} \quad (8)$$

here $\bar{B} = [\bar{b}_1, \bar{b}_2, \cdots, \bar{b}_\Xi]$ and $\bar{a} = [a_1, a_2, \cdots, a_\Xi]^T$, and the superscript "$T$" represents the transpose of matrix. The symbol "$\cdot$" in (8) represents ordinary matrix multiplication.

Inserting (8) into the second equality in (3), the power $P^{out}(\bar{J}^{sca})$ can be rewritten as the following matrix form.

$$P^{out}(\bar{a}) = \bar{a}^H \cdot \bar{\bar{P}} \cdot \bar{a} \quad (9)$$

The elements of matrix $\bar{\bar{P}}$ are $p_{\xi\zeta} = -(1/2)\langle \bar{b}_\xi, \bar{E}(\bar{b}_\zeta)\rangle_{\partial V}$ for any $\xi, \zeta = 1, 2, \cdots, \Xi$, and the matrix $\bar{\bar{P}}$ can be decomposed as follows

$$\bar{\bar{P}} = \bar{\bar{P}}_+ + j\,\bar{\bar{P}}_- \quad (10.1)$$

here

$$\bar{\bar{P}}_+ = \frac{1}{2}(\bar{\bar{P}} + \bar{\bar{P}}^H)$$

$$\bar{\bar{P}}_- = \frac{1}{2j}(\bar{\bar{P}} - \bar{\bar{P}}^H) \quad (10.2)$$

The superscript "$H$" represents the transpose conjugate of matrix. It is obvious that the matrices $\bar{\bar{P}}_+$ and $\bar{\bar{P}}_-$ are Hermitian, so the numbers $\bar{a}^H \cdot \bar{\bar{P}}_+ \cdot \bar{a}$ and $\bar{a}^H \cdot \bar{\bar{P}}_- \cdot \bar{a}$ are real for any $\bar{a}$ [16], and then

$$\bar{a}^H \cdot \bar{\bar{P}}_+ \cdot \bar{a} = P^{out}_{rad}(\bar{a}) \quad (11.1)$$

$$\bar{a}^H \cdot \bar{\bar{P}}_- \cdot \bar{a} = P^{out}_{sto}(\bar{a}) \quad (11.2)$$

and

$$Z(\bar{a}) = \begin{cases} \bar{a}^H \cdot \bar{\bar{P}} \cdot \bar{a} / \bar{a}^H \cdot \bar{\bar{J}} \cdot \bar{a} &, (\bar{a} \neq 0) \\ 0 &, (\bar{a} = 0) \end{cases} \quad (12.1)$$

$$R(\bar{a}) = \begin{cases} \bar{a}^H \cdot \bar{\bar{P}}_+ \cdot \bar{a} / \bar{a}^H \cdot \bar{\bar{J}} \cdot \bar{a} &, (\bar{a} \neq 0) \\ 0 &, (\bar{a} = 0) \end{cases} \quad (12.2)$$

$$X(\bar{a}) = \begin{cases} \bar{a}^H \cdot \bar{\bar{P}}_- \cdot \bar{a} / \bar{a}^H \cdot \bar{\bar{J}} \cdot \bar{a} &, (\bar{a} \neq 0) \\ 0 &, (\bar{a} = 0) \end{cases} \quad (12.3)$$

here $\bar{a}^H \cdot \bar{\bar{J}} \cdot \bar{a} = (1/2) \langle \bar{J}^{sca}(\bar{a}), \bar{J}^{sca}(\bar{a}) \rangle_{\partial V}$, and the elements of $\bar{\bar{J}}$ are $j_{\xi\zeta} = (1/2) \langle \bar{b}_\xi, \bar{b}_\zeta \rangle_{\partial V}$. Obviously, the matrix $\bar{\bar{P}}_+$ is either positive definite or positive semi-definite, and the matrix $\bar{\bar{J}}$ is Hermitian and positive definite.

For a certain linear PEC system, all currents which can be excited on it constitute a linear space called as current space; correspondingly, the expansion vectors constitute a vector space; for a certain $\{\bar{b}_\xi\}_{\xi=1}^\Xi$, the current space and the vector space are isomorphic, so they are not specifically distinguished from each other in the following sections. The space has many different basis sets, and every basis set can span whole space. The main destination of PEC-EMP-CMT is to select an electromagnetic power as the objective power, and then to search for a special basis set, such that the objective power corresponding to this special basis set has some desired characteristics. The elements in this special basis set are called as modal currents/vectors or characteristic currents/vectors, and the corresponding fields are called as modal fields or characteristic fields. The modal currents, vectors, and fields are collectively referred to as CMs.

In fact, the PEC-EMP-CMT is an object-oriented theory, because the selection for the objective power is relatively flexible. For example, the normalized system radiated power $R(\bar{J}^{sca})$, the normalized system stored power $X(\bar{J}^{sca})$, and the system output power $P^{out}(\bar{J}^{sca})$ are respectively selected as the objective powers in the following Secs. IV, V, and VI. However, it must be clearly pointed out that the CM sets constructed for different objective powers are not necessarily identical to each other, so the different superscripts are added to different sets to distinguish them from each other.

## IV. RADIATED POWER CHARACTERISTIC MODE (RACM) SET AND RACM-BASED MODAL EXPANSION

In fact, the normalized radiated power $R(\bar{a})$ quantitively reflects the radiation ability of the distribution of $\bar{J}^{sca}(\bar{a})$. The CM set which spans whole space and can provide the extremes of $R(\bar{a})$ is constructed in this section.

### A. Radiated power CM (RaCM) set.

The matrices $\bar{\bar{P}}_+$ and $\bar{\bar{J}}$ are Hermitian, and the $\bar{\bar{J}}$ is positive definite, so the necessary condition to achieve the extremes of $R(\bar{a})$ is following generalized characteristic equation [16].

$$\bar{\bar{P}}_+ \cdot \bar{a}_\xi^{rad} = \lambda_\xi^{rad} \bar{\bar{J}} \cdot \bar{a}_\xi^{rad} \quad (13)$$

here $\xi = 1, 2, \cdots, \Xi$, and the characteristic values are real [16].

The modal currents and fields are as follows

$$\bar{J}_\xi^{rad}(\bar{r}) = \bar{\bar{B}} \cdot \bar{a}_\xi^{rad} \quad, \quad (\bar{r} \in \partial V) \quad (14.1)$$

$$\bar{F}_\xi^{rad}(\bar{r}) = \bar{F}(\bar{J}_\xi^{rad}) \quad, \quad (\bar{r} \in \mathbb{R}^3) \quad (14.2)$$

for any $\xi = 1, 2, \cdots, \Xi$. The radiated modal powers and their normalized versions are as follows

$$P_{rad,\xi}^{out} = (\bar{a}_\xi^{rad})^H \cdot \bar{\bar{P}}_+ \cdot \bar{a}_\xi^{rad} \quad (15.1)$$

$$R_\xi^{rad} = \lambda_\xi^{rad} = \frac{(\bar{a}_\xi^{rad})^H \cdot \bar{\bar{P}}_+ \cdot \bar{a}_\xi^{rad}}{(\bar{a}_\xi^{rad})^H \cdot \bar{\bar{J}} \cdot \bar{a}_\xi^{rad}} \quad (15.2)$$

for any $\xi = 1, 2, \cdots, \Xi$.

As proven in the Appendix of this paper, the coupling radiated modal powers satisfy the following orthogonality for any $\xi, \zeta = 1, 2, \cdots, \Xi$.

$$\begin{aligned} P_{rad,\xi}^{out} \delta_{\xi\zeta} &= (\bar{a}_\xi^{rad})^H \cdot \bar{\bar{P}}_+ \cdot \bar{a}_\zeta^{rad} = \frac{1}{2} \oiint_{S_\infty} \left[ \bar{E}_\xi^{rad} \times (\bar{H}_\zeta^{rad})^* \right] \cdot d\bar{S} \\ &= \frac{1}{2\eta_0} \langle \bar{E}_\xi^{rad}, \bar{E}_\zeta^{rad} \rangle_{S_\infty} \\ &= \frac{\eta_0}{2} \langle \bar{H}_\xi^{rad}, \bar{H}_\zeta^{rad} \rangle_{S_\infty} \end{aligned} \quad (16)$$

here the $\delta_{\xi\zeta}$ is Kronecker delta symbol.

The CMs derived above are specifically called as Radiated power CMs (RaCMs) to be distinguished from the other kinds of CMs constructed in Secs. V and VI. Because $P_{rad,\xi}^{out} \geq 0$ for any $\xi = 1, 2, \cdots, \Xi$, all RaCMs can be divided into two categories, that are the radiative RaCMs corresponding to $P_{rad,\xi}^{out} > 0$ ($\xi = r_1, r_2, \cdots, r_R$) and the non-radiative RaCMs corresponding to $P_{rad,\xi}^{out} = 0$ ($\xi = n_1, n_2, \cdots, n_N$), here $\{r_1, r_2, \cdots, r_R\} \cap \{n_1, n_2, \cdots, n_N\} = \varnothing$, and $\{r_1, r_2, \cdots, r_R\} \cup \{n_1, n_2, \cdots, n_N\} = \{1, 2, \cdots, \Xi\}$.

### B. RaCM-based modal expansion.

Because of the completeness of basis function set $\{\bar{b}_\xi\}_{\xi=1}^\Xi$ and the characteristic vector set $\{\bar{a}_\xi^{rad}\}_{\xi=1}^\Xi$ [16], the scattering vector $\bar{a}^{sca}$, the scattering current $\bar{J}^{sca}$, and the scattering field $\bar{F}^{sca}$ can be expanded in terms of the RaCM set as follows

$$\bar{a}^{sca} = \sum_{\xi=1}^\Xi c_\xi^{rad} \bar{a}_\xi^{rad} \quad (17.1)$$

$$\bar{J}^{sca}(\bar{r}) = \sum_{\xi=1}^\Xi c_\xi^{rad} \bar{J}_\xi^{rad}(\bar{r}) \quad, \quad (\bar{r} \in \partial V) \quad (17.2)$$

$$\bar{F}^{sca}(\bar{r}) = \sum_{\xi=1}^\Xi c_\xi^{rad} \bar{F}_\xi^{rad}(\bar{r}) \quad, \quad (\bar{r} \in \mathbb{R}^3) \quad (17.3)$$

Based on the orthogonality in (16), the system radiated power can be expanded in terms of the radiated modal powers as

$$P_{rad}^{out} = (\bar{a}^{sca})^H \cdot \bar{\bar{P}}_+ \cdot \bar{a}^{sca} = \sum_{\xi=1}^\Xi |c_\xi^{rad}|^2 P_{rad,\xi}^{out} \quad (18)$$

*C. Expansion coefficients.*

Based on the discussions in Sec. II, the $P_{rad}^{out}(\bar{J}^{sca})$ can be written as follows

$$P_{rad}^{out}(\bar{J}^{sca}) = \text{Re}\{P^{inp}(\bar{J}^{sca})\} = \text{Re}\left\{\frac{1}{2}\langle \bar{J}^{sca}, \bar{E}^{inc}\rangle_{\partial V}\right\} \quad (19)$$

Because of the (11) and (19), the $\bar{J}^{sca} = \bar{\bar{B}} \cdot \bar{a}^{sca}$ on scatterer will make the following functional be zero and stationary [17].

$$F_1(\bar{a}^{sca}) = \text{Re}\left\{\frac{1}{2}\langle \bar{\bar{B}} \cdot \bar{a}^{sca}, \bar{E}^{inc}\rangle_{\partial V}\right\} - (\bar{a}^{sca})^H \cdot \bar{\bar{P}}_+ \cdot \bar{a}^{sca} \quad (20)$$

Inserting the (17.1) into (20) and employing the Ritz's procedure [18], the following simultaneous equations for the expansion coefficients $\{c_\xi^{rad}\}_{\xi=1}^{\Xi}$ are derived for any $\xi = 1, 2, \cdots, \Xi$.

$$\begin{aligned}
0 = &\text{Re}\left\{\frac{1}{2}\langle \bar{\bar{B}} \cdot \bar{a}_\xi^{rad}, \bar{E}^{inc}\rangle_{\partial V}\right\} \\
&- (\bar{a}_\xi^{rad})^H \cdot \bar{\bar{P}}_+ \cdot \left(\sum_{\zeta=1}^{\Xi} c_\zeta^{rad} \bar{a}_\zeta^{rad}\right) - \left(\sum_{\zeta=1}^{\Xi} c_\zeta^{rad} \bar{a}_\zeta^{rad}\right)^H \cdot \bar{\bar{P}}_+ \cdot \bar{a}_\xi^{rad}
\end{aligned} \quad (21.1)$$

$$\begin{aligned}
0 = &\text{Im}\left\{\frac{1}{2}\langle \bar{\bar{B}} \cdot \bar{a}_\xi^{rad}, \bar{E}^{inc}\rangle_{\partial V}\right\} \\
&+ j(\bar{a}_\xi^{rad})^H \cdot \bar{\bar{P}}_+ \cdot \left(\sum_{\zeta=1}^{\Xi} c_\zeta^{rad} \bar{a}_\zeta^{rad}\right) - j\left(\sum_{\zeta=1}^{\Xi} c_\zeta^{rad} \bar{a}_\zeta^{rad}\right)^H \cdot \bar{\bar{P}}_+ \cdot \bar{a}_\xi^{rad}
\end{aligned} \quad (21.2)$$

By employing the orthogonality in (16), the (21) can be rewritten as the following (22) for any $\xi = 1, 2, \cdots, \Xi$.

$$\begin{aligned}
0 = &\text{Re}\left\{\frac{1}{2}\langle \bar{J}_\xi^{rad}, \bar{E}^{inc}\rangle_{\partial V}\right\} \\
&- c_\xi^{rad}(\bar{a}_\xi^{rad})^H \cdot \bar{\bar{P}}_+ \cdot \bar{a}_\xi^{rad} - (c_\xi^{rad})^*(\bar{a}_\xi^{rad})^H \cdot \bar{\bar{P}}_+ \cdot \bar{a}_\xi^{rad}
\end{aligned} \quad (22.1)$$

$$\begin{aligned}
0 = &\text{Im}\left\{\frac{1}{2}\langle \bar{J}_\xi^{rad}, \bar{E}^{inc}\rangle_{\partial V}\right\} \\
&+ j\, c_\xi^{rad}(\bar{a}_\xi^{rad})^H \cdot \bar{\bar{P}}_+ \cdot \bar{a}_\xi^{rad} - j(c_\xi^{rad})^*(\bar{a}_\xi^{rad})^H \cdot \bar{\bar{P}}_+ \cdot \bar{a}_\xi^{rad}
\end{aligned} \quad (22.2)$$

i.e.,

$$\left[c_\xi^{rad} + (c_\xi^{rad})^*\right] \cdot P_{rad,\xi}^{out} = \text{Re}\left\{\frac{1}{2}\langle \bar{J}_\xi^{rad}, \bar{E}^{inc}\rangle_{\partial V}\right\} \quad (23.1)$$

$$\left[c_\xi^{rad} - (c_\xi^{rad})^*\right] \cdot P_{rad,\xi}^{out} = j\,\text{Im}\left\{\frac{1}{2}\langle \bar{J}_\xi^{rad}, \bar{E}^{inc}\rangle_{\partial V}\right\} \quad (23.2)$$

Because the $P_{rad,\xi}^{out}$ are real numbers, the two equations in (23) can be uniquely written as following one for any $\xi = 1, 2, \cdots, \Xi$.

$$2\, P_{rad,\xi}^{out}\, c_\xi^{rad} = \frac{1}{2}\langle \bar{J}_\xi^{rad}, \bar{E}^{inc}\rangle_{\partial V} \quad (24)$$

**1) when the matrix $\bar{\bar{P}}_+$ is positive definite**

When the matrix $\bar{\bar{P}}_+$ is positive definite, $P_{rad,\xi}^{out} > 0$ for any $\xi = 1, 2, \cdots, \Xi$, and then all $c_\xi^{rad}$ can be explicitly expressed as

$$c_\xi^{rad} = \frac{1}{2 P_{rad,\xi}^{out}} \cdot \frac{1}{2}\langle \bar{J}_\xi^{rad}, \bar{E}^{inc}\rangle_{\partial V} \quad (25)$$

The symbol "$\cdot$" in (25) is the ordinary scalar multiplication.

**2) when the matrix $\bar{\bar{P}}_+$ is positive semi-definite**

When the matrix $\bar{\bar{P}}_+$ is positive semi-definite, the radiative RaCMs corresponding to $P_{rad,\xi}^{out} > 0$ ($\xi = r_1, r_2, \cdots, r_R$) and the non-radiative RaCMs corresponding to $P_{rad,\xi}^{out} = 0$ ($\xi = n_1, n_2, \cdots, n_N$) simultaneously exist in the RaCM set. The expansion coefficients of the radiative RaCMs can also be expressed as (25), but the coefficients of the non-radiative RaCMs cannot.

In fact, the non-radiative RaCMs are just the interior resonant modes of closed PEC structures as proven in Sec. VIII-A, and the interior resonant modes should not be included in the expansion formulation for external scattering problems [13], and then the expansion coefficients of all non-radiative RaCMs are zeros for external scattering problems.

V. STORED POWER CHARACTERISTIC MODE (StoCM) SET AND StoCM-BASED MODAL EXPANSION

Similarly to the $R(\bar{a})$, the normalized stored power $X(\bar{a})$ quantitively reflects the scatterer's ability to reactively store energies. The CM set which spans whole space and can provide the extremes of $X(\bar{a})$ is constructed in this section.

*A. Stored power CM (StoCM) set.*

The matrices $\bar{\bar{P}}_-$ and $\bar{\bar{J}}$ are Hermitian, and the matrix $\bar{\bar{J}}$ is positive definite, so the necessary condition to achieve the extremes of $X(\bar{a})$ is the following characteristic equation [16].

$$\bar{\bar{P}}_- \cdot \bar{a}_\xi^{sto} = \lambda_\xi^{sto}\, \bar{\bar{J}} \cdot \bar{a}_\xi^{sto} \quad (26)$$

here $\xi = 1, 2, \cdots, \Xi$, and all characteristic values are real [16].

The modal currents and modal fields can be similarly obtained as (14). The reactively stored modal powers and their normalized versions are as follows for any $\xi = 1, 2, \cdots, \Xi$.

$$P_{sto,\xi}^{out} = (\bar{a}_\xi^{sto})^H \cdot \bar{\bar{P}}_- \cdot \bar{a}_\xi^{sto} \quad (27.1)$$

$$X_\xi^{sto} = \lambda_\xi^{sto} = \frac{(\bar{a}_\xi^{sto})^H \cdot \bar{\bar{P}}_- \cdot \bar{a}_\xi^{sto}}{(\bar{a}_\xi^{sto})^H \cdot \bar{\bar{J}} \cdot \bar{a}_\xi^{sto}} \quad (27.2)$$

By doing some necessary orthogonalizations for the degenerate modes [16], the following orthogonality holds for any $\xi, \zeta = 1, 2, \cdots, \Xi$.

$$\begin{aligned}
P_{sto,\xi}^{out} \delta_{\xi\zeta} &= (\bar{a}_\xi^{sto})^H \cdot \bar{\bar{P}}_- \cdot \bar{a}_\zeta^{sto} \\
&= 2\omega\left[\frac{1}{4}\langle \bar{H}_\xi^{sto}, \mu_0 \bar{H}_\zeta^{sto}\rangle_{\mathbb{R}^3} - \frac{1}{4}\langle \varepsilon_0 \bar{E}_\xi^{sto}, \bar{E}_\zeta^{sto}\rangle_{\mathbb{R}^3}\right]
\end{aligned} \quad (28)$$

The proof for the orthogonality (28) is similar to the proof for (16) given in Appendix, and the proof for (28) will not be repeated in this paper.

The modes derived above are specifically called as reactively Stored power CMs (StoCMs) to be distinguished from the other kinds of CMs constructed in Secs. IV and VI. All StoCMs can be divided into three categories, that are the inductive StoCMs corresponding to $P_{sto,\xi}^{out} > 0$ ( $\xi = \iota_1, \iota_2, \cdots, \iota_I$ ), the capacitive StoCMs corresponding to $P_{sto,\xi}^{out} < 0$ ( $\xi = \chi_1, \chi_2, \cdots, \chi_X$ ), and the resonant StoCMs corresponding to $P_{sto,\xi}^{out} = 0$ ( $\xi = \rho_1, \rho_2, \cdots, \rho_P$ ). Here, $\{\iota_1, \iota_2, \cdots, \iota_I\} \cup \{\chi_1, \chi_2, \cdots, \chi_X\} \cup \{\rho_1, \rho_2, \cdots, \rho_P\} = \{1, 2, \cdots, \Xi\}$; the index set $\{\iota_1, \iota_2, \cdots, \iota_I\}$, the index set $\{\chi_1, \chi_2, \cdots, \chi_X\}$, and the index set $\{\rho_1, \rho_2, \cdots, \rho_P\}$ are pairwise disjoint. In fact, the resonant StoCMs can be further divided into two categories based on their radiated powers, that are the radiative resonant StoCMs corresponding to the positive radiated powers and the non-radiative resonant StoCMs corresponding to the zero radiated powers. As proven in Sec. VIII-A, the non-radiative resonant StoCMs are just the interior resonant modes of closed PEC structures, so they can also be called as interior resonant modes. Similarly, the radiative resonant StoCMs can be also called as exterior resonant modes.

*B. StoCM-based modal expansion.*

Because of the completeness of basis function set $\{\bar{b}_\xi\}_{\xi=1}^{\Xi}$ and the characteristic vector set $\{\bar{a}_\xi^{sto}\}_{\xi=1}^{\Xi}$ [16], the scattering vector $\bar{a}^{sca}$, the scattering current $\bar{J}^{sca}$, and the scattering field $\bar{F}^{sca}$ can be expanded in terms of the StoCM set as follows

$$\bar{a}^{sca} = \sum_{\xi=1}^{\Xi} c_\xi^{sto} \bar{a}_\xi^{sto} \tag{29.1}$$

$$\bar{J}^{sca}(\bar{r}) = \sum_{\xi=1}^{\Xi} c_\xi^{sto} \bar{J}_\xi^{sto}(\bar{r}) \quad , \quad (\bar{r} \in \partial V) \tag{29.2}$$

$$\bar{F}^{sca}(\bar{r}) = \sum_{\xi=1}^{\Xi} c_\xi^{sto} \bar{F}_\xi^{sto}(\bar{r}) \quad , \quad (\bar{r} \in \mathbb{R}^3) \tag{29.3}$$

Based on the orthogonality given in (28), the system stored power can be expanded in terms of the stored modal powers as

$$P_{sto}^{out} = (\bar{a}^{sca})^H \cdot \bar{\bar{P}}_- \cdot \bar{a}^{sca} = \sum_{\xi=1}^{\Xi} |c_\xi^{sto}|^2 P_{sto,\xi}^{out} \tag{30}$$

*C. Expansion coefficients.*

Based on Sec. II, the $P_{sto}^{out}(\bar{J}^{sca})$ can be written as follows

$$P_{sto}^{out}(\bar{J}^{sca}) = \text{Im}\left\{\frac{1}{2}\langle \bar{J}^{sca}, \bar{E}^{inc} \rangle_{\partial V}\right\} \tag{31}$$

Based on the (11) and (31), the $\bar{J}^{sca} = \bar{\bar{B}} \cdot \bar{a}^{sca}$ on scatterer will make the following functional be zero and stationary.

$$F_2(\bar{a}^{sca}) = \text{Im}\left\{\frac{1}{2}\langle \bar{\bar{B}} \cdot \bar{a}^{sca}, \bar{E}^{inc} \rangle_{\partial V}\right\} - (\bar{a}^{sca})^H \cdot \bar{\bar{P}}_- \cdot \bar{a}^{sca} \tag{32}$$

Inserting the (29.1) into (32) and employing the Ritz's procedure [18], the following simultaneous equations for the expansion coefficients $\{c_\xi^{sto}\}_{\xi=1}^{\Xi}$ are derived for any $\xi = 1, 2, \cdots, \Xi$.

$$0 = \text{Im}\left\{\frac{1}{2}\langle \bar{\bar{B}} \cdot \bar{a}_\xi^{sto}, \bar{E}^{inc} \rangle_{\partial V}\right\}$$
$$- (\bar{a}_\xi^{sto})^H \cdot \bar{\bar{P}}_- \cdot \left(\sum_{\zeta=1}^{\Xi} c_\zeta^{sto} \bar{a}_\zeta^{sto}\right) - \left(\sum_{\zeta=1}^{\Xi} c_\zeta^{sto} \bar{a}_\zeta^{sto}\right)^H \cdot \bar{\bar{P}}_- \cdot \bar{a}_\xi^{sto} \tag{33.1}$$

$$0 = -\text{Re}\left\{\frac{1}{2}\langle \bar{\bar{B}} \cdot \bar{a}_\xi^{sto}, \bar{E}^{inc} \rangle_{\partial V}\right\}$$
$$+ j(\bar{a}_\xi^{sto})^H \cdot \bar{\bar{P}}_- \cdot \left(\sum_{\zeta=1}^{\Xi} c_\zeta^{sto} \bar{a}_\zeta^{sto}\right) - j\left(\sum_{\zeta=1}^{\Xi} c_\zeta^{sto} \bar{a}_\zeta^{sto}\right)^H \cdot \bar{\bar{P}}_- \cdot \bar{a}_\xi^{sto} \tag{33.2}$$

By employing the orthogonality in (28), the (33) can be rewritten as

$$\left[c_\xi^{sto} + (c_\xi^{sto})^*\right] j P_{sto,\xi}^{out} = j \text{Im}\left\{\frac{1}{2}\langle \bar{J}_\xi^{sto}, \bar{E}^{inc} \rangle_{\partial V}\right\} \tag{34.1}$$

$$\left[c_\xi^{sto} - (c_\xi^{sto})^*\right] j P_{sto,\xi}^{out} = \text{Re}\left\{\frac{1}{2}\langle \bar{J}_\xi^{sto}, \bar{E}^{inc} \rangle_{\partial V}\right\} \tag{34.2}$$

Because the $P_{sto,\xi}^{out}$ is real, the two equations in (34) can be uniquely written as the following one for any $\xi = 1, 2, \cdots, \Xi$.

$$j 2 P_{sto,\xi}^{out} c_\xi^{sto} = \frac{1}{2}\langle \bar{J}_\xi^{sto}, \bar{E}^{inc} \rangle_{\partial V} \tag{35}$$

For any non-resonant StoCM, $P_{sto,\xi}^{out} \neq 0$, and then the expansion coefficients corresponding to the non-resonant StoCMs can be explicitly expressed as the following (36) for any $\xi = \iota_1, \iota_2, \cdots, \iota_I, \chi_1, \chi_2, \cdots, \chi_X$.

$$c_\xi^{sto} = \frac{1}{j 2 P_{sto,\xi}^{out}} \cdot \frac{1}{2}\langle \bar{J}_\xi^{sto}, \bar{E}^{inc} \rangle_{\partial V} \tag{36}$$

For the radiative resonant StoCMs (exterior resonant modes), there are not the concise expressions for their expansion coefficients. For the external scattering problems, the non-radiative resonant StoCMs (interior resonant modes) should not be included in the expansion formulation [13], so their expansion coefficients are zeros.

## VI. OUTPUT POWER CHARACTERISTIC MODE (OutCM) SET AND OutCM-BASED MODAL EXPANSION

The CM set which satisfies the output power orthogonality in (45) has some very good properties, and it is constructed in this section, and it is specifically called as the Output power CM (OutCM) set to be distinguished from the other kinds of CM sets constructed in the Secs. IV and V. It must be clearly pointed out here that the OutCM set has the same physical essence as the CM set derived from PEC-SEFIE-CMT [2]-[3]. However, the OutCM set is more advantageous in some aspects, as discussed in the Sec. VIII-D.

*A. Output power CM (OutCM) set.*

The matrices $\bar{\bar{P}}_+$ and $\bar{\bar{P}}_-$ are the functions of frequency $f$, so they are denoted as $\bar{\bar{P}}_+(f)$ and $\bar{\bar{P}}_-(f)$ for the convenience of following discussions. The procedures to construct OutCM set will be provided by separately focusing on two different cases: case I, there doesn't exist any non-radiative mode at frequency $f$, and then the $\bar{\bar{P}}_+(f)$ is positive definite; case II, there exist some non-radiative modes at frequency $f_0$, and then the $\bar{\bar{P}}_+(f_0)$ is positive semi-definite.

**To construct the OutCM set, when the matrix $\bar{\bar{P}}_+(f)$ is positive definite.**

Both $\bar{\bar{P}}_+(f)$ and $\bar{\bar{P}}_-(f)$ are Hermitian, and the $\bar{\bar{P}}_+(f)$ is positive definite, so there is a vector set $\{\bar{a}_\xi^{out}\}_{\xi=1}^\Xi$, such that [16]

$$\left[\bar{a}_\xi^{out}(f)\right]^H \cdot \bar{\bar{P}}_+(f) \cdot \bar{a}_\zeta^{out}(f) = P_{\xi,rad}^{out}(f)\delta_{\xi\zeta} \quad (37.1)$$

$$\left[\bar{a}_\xi^{out}(f)\right]^H \cdot \bar{\bar{P}}_-(f) \cdot \bar{a}_\zeta^{out}(f) = P_{\xi,sto}^{out}(f)\delta_{\xi\zeta} \quad (37.2)$$

for any $\xi,\zeta = 1,2,\cdots,\Xi$, and then

$$\left[\bar{a}_\xi^{out}(f)\right]^H \cdot \bar{\bar{P}}(f) \cdot \bar{a}_\zeta^{out}(f) = P_\xi^{out}(f)\delta_{\xi\zeta} \quad (38)$$

for any $\xi,\zeta = 1,2,\cdots,\Xi$, here

$$P_\xi^{out}(f) = P_{\xi,rad}^{out}(f) + j P_{\xi,sto}^{out}(f) \quad (39)$$

for any $\xi = 1,2,\cdots,\Xi$.

The characteristic vector set $\{\bar{a}_\xi^{out}\}_{\xi=1}^\Xi$ can be obtained by solving the following (40) and doing some necessary orthogonalizations for the degenerate modes [16].

$$\bar{\bar{P}}_-(f) \cdot \bar{a}_\xi^{out}(f) = \lambda_\xi^{out}(f) \bar{\bar{P}}_+(f) \cdot \bar{a}_\xi^{out}(f) \quad (40)$$

here the $\lambda_\xi^{out}$ is real [16]. The modal currents and modal fields can be similarly expressed as the (14) for any $\xi = 1,2,\cdots,\Xi$.

The modal output power is given in (39), and its normalized version (modal output impedance) is as follows

$$Z_\xi^{out}(f) = R_\xi^{out}(f) + j X_\xi^{out}(f)$$
$$= \frac{\left[\bar{a}_\xi^{out}(f)\right]^H \cdot \bar{\bar{P}}(f) \cdot \bar{a}_\xi^{out}(f)}{\left[\bar{a}_\xi^{out}(f)\right]^H \cdot \bar{\bar{J}}(f) \cdot \bar{a}_\xi^{out}(f)} \quad (41.1)$$

here

$$R_\xi^{out}(f) = \frac{\left[\bar{a}_\xi^{out}(f)\right]^H \cdot \bar{\bar{P}}_+(f) \cdot \bar{a}_\xi^{out}(f)}{\left[\bar{a}_\xi^{out}(f)\right]^H \cdot \bar{\bar{J}}(f) \cdot \bar{a}_\xi^{out}(f)} \quad (41.2)$$

$$X_\xi^{out}(f) = \frac{\left[\bar{a}_\xi^{out}(f)\right]^H \cdot \bar{\bar{P}}_-(f) \cdot \bar{a}_\xi^{out}(f)}{\left[\bar{a}_\xi^{out}(f)\right]^H \cdot \bar{\bar{J}}(f) \cdot \bar{a}_\xi^{out}(f)} \quad (41.3)$$

The $P_{\xi,rad}^{out}(f)$ and $R_\xi^{out}(f)$ are respectively called as modal radiated power and modal radiated/output resistance; The $P_{\xi,sto}^{out}(f)$ and $X_\xi^{out}(f)$ are called as modal stored power and modal stored/output reactance respectively.

In fact, it is obvious that the characteristic values $\lambda_\xi^{out}(f)$ derived from (40) satisfy the following relation (42) for any $\xi = 1,2,\cdots,\Xi$

$$\lambda_\xi^{out}(f) = \frac{P_{\xi,sto}^{out}(f)}{P_{\xi,rad}^{out}(f)} = \frac{X_\xi^{out}(f)}{R_\xi^{out}(f)} \quad (42)$$

and the characteristic equation (40) is just the necessary condition to achieve the extreme of following functional [16].

$$F_3(\bar{a}) = \frac{P_{sto}^{out}(\bar{a})}{P_{rad}^{out}(\bar{a})} = \frac{X^{out}(\bar{a})}{R^{out}(\bar{a})} = \frac{\bar{a}^H \cdot \bar{\bar{P}}_- \cdot \bar{a}}{\bar{a}^H \cdot \bar{\bar{P}}_+ \cdot \bar{a}} \quad (43)$$

**To construct the OutCM set, when the matrix $\bar{\bar{P}}_+(f_0)$ is positive semi-definite.**

When the matrix $\bar{\bar{P}}_+(f_0)$ is positive semi-definite, there doesn't exist a ready-made mathematical theorem to support that the vector set $\{\bar{a}_\xi^{out}\}_{\xi=1}^\Xi$ satisfying (37)-(38) can be obtained by solving (40). In this paper, a limitation method is employed to derive the $\{\bar{a}_\xi^{out}\}_{\xi=1}^\Xi$ at $f_0$, and it includes two core steps: step I, to find the $f_0$ at which the $\bar{\bar{P}}_+(f_0)$ is positive semi-definite; step II, to construct the OutCM set at $f_0$.

Step I, to find the $f_0$

Due to the limitation viewpoint, if the $\bar{\bar{P}}_+(f_0)$ is positive semi-definite, some radiative modes at $f$ will approach to the non-radiative modes at $f_0$, when $f \to f_0$. Because the non-radiative mode is also the resonant mode as proven in Sec. VIII-A, its $Z_\xi^{out}(f_0)$ equals to zero, and then the limitation of corresponding characteristic value $\lambda_\xi^{out}(f) = X_\xi^{out}(f)/R_\xi^{out}(f)$ is an indeterminate form during $f \to f_0$ [19], so it is sometimes difficult to efficiently recognize the $f_0$ only based on the limitation $\lim_{f \to f_0} \lambda_\xi^{out}(f)$. In fact, the $Z_\xi^{out}(f)$, which equals to $R_\xi^{out}(f) + j X_\xi^{out}(f)$, is a more complete description for the modal characteristic to utilize various energies than the $\lambda_\xi^{out}(f)$, because the $\lambda_\xi^{out}(f)$ has only ability to depict the ratio of the $X_\xi^{out}(f)$ to the $R_\xi^{out}(f)$, but the $X_\xi^{out}(f)$ and $R_\xi^{out}(f)$ themselves cannot be explicitly reflected by the $\lambda_\xi^{out}(f)$. Based on above observations, a modal impedance based method to search for the frequency $f_0$ is developed here.

The $R_\xi^{out}(f)$ and $X_\xi^{out}(f)$ are the one-variable real functions about $f$, and their function curves can be easily obtained by using frequency sweep technique. The $R_\xi^{out}(f)$ curve will pass through the point $(f_0,0)$, if and only if the mode $\xi$ doesn't radiate at the frequency $f_0$.

Step II, to construct the OutCM set at $f_0$

When the $f_0$ is found out, the modal vectors, currents, fields, powers, and impedances at $f_0$ can be obtained as the following limitations for any $\xi = 1,2,\cdots,\Xi$.

$$\bar{a}_\xi^{out}(f_0) = \lim_{f \to f_0} \bar{a}_\xi^{out}(f) \quad (44.1)$$

$$\bar{J}_\xi^{out}(f_0) = \lim_{f \to f_0} \bar{J}_\xi^{out}(f) \quad (44.2)$$

$$\bar{F}_\xi^{out}(f_0) = \lim_{f \to f_0} \bar{F}_\xi^{out}(f) \quad (44.3)$$

$$P_\xi^{out}(f_0) = \lim_{f \to f_0} P_\xi^{out}(f) \quad (44.4)$$

$$Z_\xi^{out}(f_0) = \lim_{f \to f_0} Z_\xi^{out}(f) \quad (44.5)$$

**Summarizing both the positive definite $\bar{\bar{P}}_+(f)$ and positive semi-definite $\bar{\bar{P}}_+(f_0)$ cases.**

Based on the above discussions, the coupling modal output powers for any radiative modes $r_1, r_2, \cdots, r_R$ and non-radiative modes $n_1, n_2, \cdots, n_N$ satisfy the following output power orthogonality.

$$P_\xi^{out} \delta_{\xi\zeta} = (\bar{a}_\xi^{out})^H \cdot \bar{\bar{P}} \cdot \bar{a}_\zeta^{out} = -\frac{1}{2} \langle \bar{J}_\xi^{out}, \bar{E}_\zeta^{out} \rangle_{\partial V} \quad (45.1)$$

and

$$P_{\xi,rad}^{out} \delta_{\xi\zeta} = (\bar{a}_\xi^{out})^H \cdot \bar{\bar{P}}_+ \cdot \bar{a}_\zeta^{out} = \frac{1}{2} \oiint_{S_\infty} [\bar{E}_\xi^{out} \times (\bar{H}_\xi^{out})^*] \cdot d\bar{S}$$
$$= \frac{1}{2\eta_0} \langle \bar{E}_\xi^{out}, \bar{E}_\zeta^{out} \rangle_{S_\infty} \quad (45.2)$$
$$= \frac{\eta_0}{2} \langle \bar{H}_\xi^{out}, \bar{H}_\zeta^{out} \rangle_{S_\infty}$$

$$P_{\xi,sto}^{out} \delta_{\xi\zeta} = (\bar{a}_\xi^{out})^H \cdot \bar{\bar{P}}_- \cdot \bar{a}_\zeta^{out}$$
$$= 2\omega \left[ \frac{1}{4} \langle \bar{H}_\xi^{out}, \mu_0 \bar{H}_\zeta^{out} \rangle_{\mathbb{R}^3} - \frac{1}{4} \langle \varepsilon_0 \bar{E}_\xi^{out}, \bar{E}_\zeta^{out} \rangle_{\mathbb{R}^3} \right] \quad (45.3)$$

here

$$P_\xi^{out} = \begin{cases} P_{\xi,rad}^{out} + j P_{\xi,sto}^{out} &, (\xi = r_1, r_2, \cdots, r_R) \\ 0 &, (\xi = n_1, n_2, \cdots, n_N) \end{cases} \quad (46)$$

In (46), the conclusion that the non-radiative modes must be the resonant modes has been used. The proof for (45) is similar to the proof for (16) given in Appendix.

That the symbols "$P_{\xi,rad}^{out}$" and "$P_{\xi,sto}^{out}$" are not respectively written as the "$P_{rad,\xi}^{out}$" in Sec. IV and the "$P_{sto,\xi}^{out}$" in Sec. V is to emphasize that the "$P_{\xi,rad}^{out}$" and "$P_{\xi,sto}^{out}$" are the real and imaginary parts of the modal output power $P_\xi^{out}$, whereas the "$P_{rad,\xi}^{out}$" and "$P_{sto,\xi}^{out}$" are just the modal powers themselves derived by optimizing $R(\bar{a})$ and $X(\bar{a})$. In fact, this is just the reason why the $P_{\xi,rad}^{out}$ is called as "modal radiated power" in this section, instead of being called as the "radiated modal power" like the $P_{rad,\xi}^{out}$ in Sec. IV. Other terms can be similarly explained.

*B. Modal expansion.*

Similarly to the Secs. IV and V, the scattering vector $\bar{a}^{sca}$, the scattering current $\bar{J}^{sca}$, and the scattering field $\bar{F}^{sca}$ can be expanded in terms of the OutCM set as follows

$$\bar{a}^{sca} = \sum_{\xi=1}^{\Xi} c_\xi^{out} \bar{a}_\xi^{out} \quad (47.1)$$

$$\bar{J}^{sca}(\bar{r}) = \sum_{\xi=1}^{\Xi} c_\xi^{out} \bar{J}_\xi^{out}(\bar{r}) \, , \, (\bar{r} \in \partial V) \quad (47.2)$$

$$\bar{F}^{sca}(\bar{r}) = \sum_{\xi=1}^{\Xi} c_\xi^{out} \bar{F}_\xi^{out}(\bar{r}) \, , \, (\bar{r} \in \mathbb{R}^3) \quad (47.3)$$

Based on the power orthogonality (45), various system powers can be expanded as follows

$$P^{out} = (\bar{a}^{sca})^H \cdot \bar{\bar{P}} \cdot \bar{a}^{sca} = \sum_{\xi=1}^{\Xi} |c_\xi^{out}|^2 P_\xi^{out} \quad (48.1)$$

$$P_{rad}^{out} = (\bar{a}^{sca})^H \cdot \bar{\bar{P}}_+ \cdot \bar{a}^{sca} = \sum_{\xi=1}^{\Xi} |c_\xi^{out}|^2 P_{\xi,rad}^{out} \quad (48.2)$$

$$P_{sto}^{out} = (\bar{a}^{sca})^H \cdot \bar{\bar{P}}_- \cdot \bar{a}^{sca} = \sum_{\xi=1}^{\Xi} |c_\xi^{out}|^2 P_{\xi,sto}^{out} \quad (48.3)$$

*C. Expansion Coefficients.*

Based on the discussions in Sec. II, the $\bar{J}^{sca}$ on scatterer will make the following functional be zero and stationary.

$$F_4(\bar{J}^{sca}) = \frac{1}{2} \langle \bar{J}^{sca}, \bar{E}^{inc} \rangle_{\partial V} - \left[ -\frac{1}{2} \langle \bar{J}^{sca}, \bar{E}(\bar{J}^{sca}) \rangle_{\partial V} \right] \quad (49)$$

By inserting the (47.2) into (49) and employing the Ritz's procedure [18], the following simultaneous equations for the expansion coefficients $\{c_\xi^{out}\}_{\xi=1}^{\Xi}$ are derived for any $\xi = 1, 2, \cdots, \Xi$.

$$-\frac{1}{2} \langle \bar{J}_\xi^{out}, \bar{E}^{inc} \rangle_{\partial V} = \frac{1}{2} \langle \bar{J}_\xi^{out}, \sum_{\zeta=1}^{\Xi} c_\zeta^{out} \bar{E}_\zeta^{out} \rangle_{\partial V} + \frac{1}{2} \langle \sum_{\zeta=1}^{\Xi} c_\zeta^{out} \bar{J}_\zeta^{out}, \bar{E}_\xi^{out} \rangle_{\partial V} \quad (50.1)$$

$$\frac{1}{2} \langle \bar{J}_\xi^{out}, \bar{E}^{inc} \rangle_{\partial V} = -\frac{1}{2} \langle \bar{J}_\xi^{out}, \sum_{\zeta=1}^{\Xi} c_\zeta^{out} \bar{E}_\zeta^{out} \rangle_{\partial V} + \frac{1}{2} \langle \sum_{\zeta=1}^{\Xi} c_\zeta^{out} \bar{J}_\zeta^{out}, \bar{E}_\xi^{out} \rangle_{\partial V} \quad (50.2)$$

By employing the orthogonality in (45.1), the (50) becomes the following (51).

$$c_\xi^{out} P_\xi^{out} = \frac{1}{2} \langle \bar{J}_\xi^{out}, \bar{E}^{inc} \rangle_{\partial V} \quad (51.1)$$

$$(c_\xi^{out})^* P_\xi^{out} = 0 \quad (51.2)$$

for any $\xi = 1, 2, \cdots, \Xi$.

**1) when the matrix $\bar{\bar{P}}_+$ is positive definite**

When the matrix $\bar{\bar{P}}_+$ is positive definite, $(\bar{a}_\xi^{out})^H \cdot \bar{\bar{P}}_+ \cdot \bar{a}_\xi^{out} > 0$ for any $\xi = 1, 2, \cdots, \Xi$, and then $P_\xi^{out} \neq 0$ for any $\xi = 1, 2, \cdots, \Xi$. Based on this, by solving (51) the expansion coefficients $\{c_\xi^{out}\}_{\xi=1}^{\Xi}$ can be expressed as following (52) for any $\xi = 1, 2, \cdots, \Xi$.

$$c_\xi^{out} = \frac{1}{P_\xi^{out}} \cdot \frac{1}{2} \langle \bar{J}_\xi^{out}, \bar{E}^{inc} \rangle_{\partial V} \quad (52.1)$$

$$c_\xi^{out} = 0 \quad (52.2)$$

The symbol "$\cdot$" in (52.1) is the ordinary scalar multiplication. It is obvious that the (52.2) is the non-physical solution for the non-zero external excitation $\bar{E}^{inc}$, so only the (52.1) is the desired physical solution.

**2) when the matrix $\bar{\bar{P}}_+$ is positive semi-definite**

When the matrix $\bar{\bar{P}}_+$ is positive semi-definite, the radiative OutCMs corresponding to $P_\xi^{out} \neq 0$ ($\xi = r_1, r_2, \cdots, r_R$) and the non-radiative OutCMs corresponding to $P_\xi^{out} = 0$ ($\xi = n_1, n_2, \cdots, n_N$) exist simultaneously. The radiative OutCMs satisfy (51.1), and the non-radiative OutCMs satisfy (51.2). However, only based on the (51.2), the expansion coefficients for the non-radiative

OutCMs cannot be determined, because all non-radiative OutCMs (that are the interior resonant modes) belong to the zero space of EFIE operator [13].

In fact, for the external scattering problem of the PEC structures whose boundaries are closed surfaces, the interior resonant modes should not be included in the modal expansion formulation, so the expansion coefficients of all non-radiative OutCMs are zeros for the external scattering problem.

### VII. THE CHARACTERISTIC AND NON-CHARACTERISTIC QUANTITIES OF OUTCMS

In Sec. VI-B, the system electromagnetic quantities (such as $\bar{J}^{sca}$, $\bar{F}^{sca}$, and $P^{out}$) are expanded in terms of series of modal components (such as $c_\xi^{out} \bar{J}_\xi^{out}$, $c_\xi^{out} \bar{F}_\xi^{out}$, and $|c_\xi^{out}|^2 P_\xi^{out}$), and to consider of the weight of every modal component in whole expansion formulation is important. However, the components are either the complex vectors (such as $c_\xi^{out} \bar{J}_\xi^{out}$ and $c_\xi^{out} \bar{F}_\xi^{out}$) or the complex scalars (such as $|c_\xi^{out}|^2 P_\xi^{out}$), so the modal weight cannot be directly derived from the modal components themselves, and then to establish an appropriate mapping from the modal index set $\{\xi\}_{\xi=1}^{\Xi}$ to a real number set is necessary.

In this section, only the OutCM-based modal expansion for the external scattering problem is considered.

#### A. An illuminating example.

Let us consider the following example at first.

$$20\,\hat{x} + 20\,\hat{y} = \left[1 \cdot (10\,\hat{x} + 10\,\hat{y})\right] + \left[10 \cdot (1\,\hat{x} + 1\,\hat{y})\right] \quad (53)$$

Obviously, using the following mapping (54.1) to depict the weights of term $1 \cdot (10\,\hat{x}+10\,\hat{y})$ and term $10 \cdot (1\,\hat{x}+1\,\hat{y})$ is unreasonable.

$$\{1 \cdot (10\,\hat{x} + 10\,\hat{y})\ ,\ 10 \cdot (1\,\hat{x} + 1\,\hat{y})\} \leftrightarrow \{1\ ,\ 10\} \quad (54.1)$$

here the symbol $\{a,b\}$ represents the ordered array of numbers $a$ and $b$. In fact, a reasonable mapping is as follows

$$\{1 \cdot (10\hat{x}+10\hat{y})\ ,\ 10 \cdot (1\hat{x}+1\hat{y})\} \leftrightarrow \{1 \cdot \sqrt{10^2+10^2}\ ,\ 10 \cdot \sqrt{1^2+1^2}\} \quad (54.2)$$

because the (53) can be equivalently rewritten as the following (55).

$$20\hat{x}+20\hat{y} = \left[1 \cdot \sqrt{10^2+10^2} \cdot \frac{(10\hat{x}+10\hat{y})}{\sqrt{10^2+10^2}}\right] + \left[10 \cdot \sqrt{1^2+1^2} \cdot \frac{(1\hat{x}+1\hat{y})}{\sqrt{1^2+1^2}}\right] \quad (55)$$

By comparing the (53) with (55), it is easy to find out that the essential reason to lead to the inappropriate mapping (54.1) is that the terms $10\hat{x}+10\hat{y}$ and $1\,\hat{x}+1\,\hat{y}$ in (53) are not normalized.

#### B. Modal normalization.

Based on the example in Sec. VII-A, to normalize the OutCMs is necessary for establishing the appropriate mapping from the $\{\xi\}_{\xi=1}^{\Xi}$ to a real number set whose elements quantitively depict the modal weights in whole modal expansion formulation.

In this paper, the modal currents are normalized as follows

$$\tilde{\bar{J}}_\xi^{out}(r) \triangleq \frac{\bar{J}_\xi^{out}(r)}{\sqrt{(1/2)\langle \bar{J}_\xi^{out}, \bar{J}_\xi^{out}\rangle_{\partial V}}} \quad ,\quad (\bar{r}\in \partial V) \quad (56)$$

and then the modal fields and modal output powers are automatically normalized as follows

$$\tilde{\bar{F}}_\xi^{out}(r) = \frac{\bar{F}_\xi^{out}(\bar{r})}{\sqrt{(1/2)\langle \bar{J}_\xi^{out}, \bar{J}_\xi^{out}\rangle_{\partial V}}} \quad ,\quad (\bar{r}\in \mathbb{R}^3) \quad (57)$$

$$\tilde{P}_\xi^{out} = \frac{P_\xi^{out}}{(1/2)\langle \bar{J}_\xi^{out}, \bar{J}_\xi^{out}\rangle_{\partial V}} \quad (58)$$

and then the expansion coefficient (52.1) automatically becomes the following version for any radiative OutCM.

$$\tilde{c}_\xi^{out} = \frac{1}{\tilde{P}_\xi^{out}} \cdot \frac{1}{2}\left\langle \tilde{\bar{J}}_\xi^{out}, \bar{E}^{inc}\right\rangle_{\partial V} \quad (59)$$

Inserting the (58) and (59) into the OutCM-based modal expansion formulation for the $P^{out}$ in (48.1), the following equivalent version of (48.1) is obtained.

$$P^{inp} = P^{out} = \sum_{\xi=1}^{\Xi}\left|\tilde{c}_\xi^{out}\right|^2 \tilde{P}_\xi^{out}$$
$$= \sum_{\xi=1}^{\Xi} \frac{1}{\left|\tilde{P}_\xi^{out}\right|}\cdot\left|\frac{1}{2}\left\langle \tilde{\bar{J}}_\xi^{out}, \bar{E}^{inc}\right\rangle_{\partial V}\right|^2 \cdot \frac{\tilde{P}_\xi^{out}}{\left|\tilde{P}_\xi^{out}\right|} \quad (60)$$

Obviously, the magnitude of term $\tilde{P}_\xi^{out}/|\tilde{P}_\xi^{out}|$ is unit just like the terms $(10\,\hat{x}+10\,\hat{y})/\sqrt{10^2+10^2}$ and $(1\,\hat{x}+1\,\hat{y})/\sqrt{1^2+1^2}$ in (55), and then two mappings can be established as the following (61) just like the (54.2).

$$\{\xi\}_{\xi=1}^{\Xi} \leftrightarrow \left\{\text{SMS}_\xi^{sys,tot}\right\}_{\xi=1}^{\Xi} \quad (61.1)$$

$$\{\xi\}_{\xi=1}^{\Xi} \leftrightarrow \left\{\text{GMS}_\xi^{sys,tot}\right\}_{\xi=1}^{\Xi} \quad (61.2)$$

here

$$\text{SMS}_\xi^{sys,tot} \triangleq \frac{1}{\left|\tilde{P}_\xi^{out}\right|}\cdot\left|\frac{1}{2}\left\langle \tilde{\bar{J}}_\xi^{out}, \bar{E}^{inc}\right\rangle_{\partial V}\right|^2 \quad (62.1)$$

$$\text{GMS}_\xi^{sys,tot} \triangleq \frac{1}{\left|\tilde{P}_\xi^{out}\right|} \quad (62.2)$$

#### C. The generalized Modal Significance (MS) for system total power.

Based on the (60)-(62), it is easily found out that:

(1) When a specific field $\bar{F}^{inc}$ incidents on scatterer, only the first several terms, whose $\text{SMS}_\xi^{sys,tot}$ are relatively large, are necessary to be included in the truncated modal expansion formulation.

(2) Generally speaking, when the sufficient terms, which have relatively large $\text{GMS}_\xi^{sys,tot}$, are included in the modal expansion formulation, the truncated expansion formulation can basically coincide with the full-wave solution for any external excitation. However, it must be clearly pointed out that this conclusion is not always right. For example, when the expansion formulation only includes $N$ terms, and the external excitation satisfies that $\bar{F}^{inc} = -\bar{F}_{N+M}^{out}$ on $\partial V$ (here $M > 0$), the truncated expansion formulation cannot coincide with the full-wave solution, because only the ($N+M$)-th modal component is excited, whereas this component is not included in the truncated expansion formulation.

Based on the discussions in above (1) and (2), it is obvious that the $\text{SMS}_\xi^{sys,tot}$ and $\text{GMS}_\xi^{sys,tot}$ quantitively depict the modal weight in whole modal expansion formulation, so the $\text{SMS}_\xi^{sys,tot}$ is called as "the Modal Significance (MS) corresponding to the specific excitation" or simply called as "Special MS (SMS)", and the $\text{GMS}_\xi^{sys,tot}$ is called as "the MS corresponding to general excitation" or simply called as "General MS (GMS)". The SMS and GMS are collectively referred to as "the generalized MS for system total power", and this is the reason why the superscripts "$sys, tot$" are used in them.

*D. The generalized Modal Significance (MS) for system active and reactive powers.*

The real and imaginary parts of power $P^{out} = P^{inp}$ is the active power $P_{act}^{inp}$ and the reactive power $P_{react}^{inp}$. Because $\left|\tilde{c}_\xi^{out}\right|^2$ in (60) is real, the following expansions can be derived.

$$\begin{aligned}
P_{act}^{inp} = P_{rad}^{out} &= \sum_{\xi=1}^{\Xi} \left|\tilde{c}_\xi^{out}\right|^2 \text{Re}\{\tilde{P}_\xi^{out}\} \\
&= \sum_{\xi=1}^{\Xi} \frac{1}{\left|\tilde{P}_\xi^{out}\right|} \cdot \left|\frac{1}{2}\left\langle \bar{\tilde{J}}_\xi^{out}, \bar{E}^{inc}\right\rangle_{\partial V}\right|^2 \cdot \frac{\text{Re}\{\tilde{P}_\xi^{out}\}}{\left|\tilde{P}_\xi^{out}\right|}
\end{aligned} \quad (63.1)$$

$$\begin{aligned}
P_{react}^{inp} = P_{sto}^{out} &= \sum_{\xi=1}^{\Xi} \left|\tilde{c}_\xi^{out}\right|^2 \text{Im}\{\tilde{P}_\xi^{out}\} \\
&= \sum_{\xi=1}^{\Xi} \frac{1}{\left|\tilde{P}_\xi^{out}\right|} \cdot \left|\frac{1}{2}\left\langle \bar{\tilde{J}}_\xi^{out}, \bar{E}^{inc}\right\rangle_{\partial V}\right|^2 \cdot \frac{\text{Im}\{\tilde{P}_\xi^{out}\}}{\left|\tilde{P}_\xi^{out}\right|}
\end{aligned} \quad (63.2)$$

Similarly to the generalized MS for system total power, the following "generalized MS for system active and reactive powers" are introduced to quantitively depict the modal weights in whole system active and reactive powers.

$$\text{SMS}_\xi^{sys,act} \triangleq \text{SMS}_\xi^{sys,tot} \cdot \frac{\text{Re}\{\tilde{P}_\xi^{out}\}}{\left|\tilde{P}_\xi^{out}\right|} = \frac{1}{\left|\tilde{P}_\xi^{out}\right|^2} \cdot \left|\frac{1}{2}\left\langle \bar{\tilde{J}}_\xi^{out}, \bar{E}^{inc}\right\rangle_{\partial V}\right|^2 \cdot \text{Re}\{\tilde{P}_\xi^{out}\} \quad (64.1)$$

$$\text{GMS}_\xi^{sys,act} \triangleq \text{GMS}_\xi^{sys,tot} \cdot \frac{\text{Re}\{\tilde{P}_\xi^{out}\}}{\left|\tilde{P}_\xi^{out}\right|} = \frac{1}{\left|\tilde{P}_\xi^{out}\right|^2} \cdot \text{Re}\{\tilde{P}_\xi^{out}\} \quad (64.2)$$

for the active power, and

$$\text{SMS}_\xi^{sys,react} \triangleq \text{SMS}_\xi^{sys,tot} \cdot \frac{\text{Im}\{\tilde{P}_\xi^{out}\}}{\left|\tilde{P}_\xi^{out}\right|} = \frac{1}{\left|\tilde{P}_\xi^{out}\right|^2} \cdot \left|\frac{1}{2}\left\langle \bar{\tilde{J}}_\xi^{out}, \bar{E}^{inc}\right\rangle_{\partial V}\right|^2 \cdot \text{Im}\{\tilde{P}_\xi^{out}\} \quad (65.1)$$

$$\text{GMS}_\xi^{sys,react} \triangleq \text{GMS}_\xi^{sys,tot} \cdot \frac{\text{Im}\{\tilde{P}_\xi^{out}\}}{\left|\tilde{P}_\xi^{out}\right|} = \frac{1}{\left|\tilde{P}_\xi^{out}\right|^2} \cdot \text{Im}\{\tilde{P}_\xi^{out}\} \quad (65.2)$$

for the reactive power.

The reason to use the superscript "$sys$" in the $\text{SMS}_\xi^{sys,act/react}$ and $\text{GMS}_\xi^{sys,act/react}$ is the same as the reason to use the superscript "$sys$" in $\text{SMS}_\xi^{sys,tot}$ and $\text{GMS}_\xi^{sys,tot}$. For the PEC systems, the radiated power is just the active part of input power as illustrated in (5), however this conclusion doesn't hold for the material scatterer case [15]. To maintain the consistency with the ElectroMagentic-Power-based CMT for Material bodies (Mat-EMP-CMT) [20], the $\text{SMS}_\xi^{sys,act/react}$ and $\text{GMS}_\xi^{sys,act/react}$ are not respectively called as the SMS for system radiated/stored power and the GMS for system radiated/stored power, and this is just the reason to use the superscripts "$act/react$" instead of using the "$rad/sto$".

*E. The modal characteristic to allocate active and reactive powers and the modal ability to couple energy from external excitation.*

From the above discussions, it is obvious that the modal ability to transform the energy provided by external excitation to a part of the system active and reactive powers depends on the following three aspects:

(1) The modal ability to contribute system total power is quantitively depicted by the $\text{GMS}_\xi^{sys,tot}$ defined in (62.2).

(2) The modal characteristic to allocate the total modal output power to its active and reactive parts is quantitively depicted by "the Modal characteristic to Allocate modal Output Power (MAOP)" defined as the following (66).

$$\text{MAOP}_\xi^{mod,act} \triangleq \frac{\text{Re}\{\tilde{P}_\xi^{out}\}}{\left|\tilde{P}_\xi^{out}\right|} \quad (66.1)$$

$$\text{MAOP}_\xi^{mod,react} \triangleq \frac{\text{Im}\{\tilde{P}_\xi^{out}\}}{\left|\tilde{P}_\xi^{out}\right|} \quad (66.2)$$

(3) The modal ability to couple energy from external excitation is quantitively depicted by "the Modal Ability to Couple Excitation (MACE)" defined as the following (67).

$$\text{MACE}_\xi^{mod} \triangleq \left|\frac{1}{2}\left\langle \bar{\tilde{J}}_\xi^{out}, \bar{E}^{inc}\right\rangle_{\partial V}\right|^2 \quad (67)$$

The superscripts "$mod$" in (66) and (67) are to emphasize that to introduce the MAOP and MACE is to depict the characteristic of the mode itself, but not to depict the modal weight in whole system power.

*F. The characteristic and non-characteristic quantities.*

Obviously, the $\text{GMS}_\xi^{sys,tot}$, $\text{GMS}_\xi^{sys,act/react}$, and $\text{MAOP}_\xi^{mod,act/react}$ are independent of the specific external excitation, so they are collectively referred to as "modal characteristic quantities" (or simply as "characteristic quantities"). However, the $\text{SMS}_\xi^{sys,tot}$, $\text{SMS}_\xi^{sys,act/react}$, and $\text{MACE}_\xi^{mod}$ depend on the specific external excitation, so they are collectively referred to as "modal non-characteristic quantities" (or simply as "non-characteristic quantities"). The characteristic quantities and non-characteristic quantities are collectively referred to as "modal quantities".

For the PEC systems, the $\tilde{P}^{out}_{\xi,rad}$ and $\tilde{P}^{out}_{\xi,sto}$ respectively correspond to the modal far and near fields. Based on this, the following conclusions can be obtained.

(1) The generalized MS for system active and reactive powers defined in (64) and (65) quantitively depict the weights of the modal far and near fields in whole system far and near fields.

(2) The MAOP in (66) quantitively depicts the modal characteristic to allocate the modal field to its far and near field zones.

(3) To compare the MACE with the GMS$^{sys,tot}$ (GMS$^{sys,act}$, and GMS$^{sys,react}$) is an efficient method to judge if the specific excitation is suitable for exciting the objective PEC system (the far field modes, and the near field modes).

## VIII. Discussions

In this section, some necessary discussions related to the theory developed in this paper are provided.

### A. The discussions for non-radiative CMs.

The following discussions in this subsection are suitable for the any kind of CM set developed in Secs. IV, V, and VI.

When the matrix $\bar{\bar{P}}_+$ is positive semi-definite, $\bar{a}^H \cdot \bar{\bar{P}}_+ \cdot \bar{a} = 0$ if and only if $\bar{\bar{P}}_+ \cdot \bar{a} = 0$ [16], and then $\bar{a}_1^H \cdot \bar{\bar{P}}_+ \cdot \bar{a}_1 = 0$ and $\bar{a}_2^H \cdot \bar{\bar{P}}_+ \cdot \bar{a}_2 = 0$ imply that $(\alpha \bar{a}_1 + \beta \bar{a}_2)^H \cdot \bar{\bar{P}}_+ \cdot (\alpha \bar{a}_1 + \beta \bar{a}_2) = 0$ for any complex numbers $\alpha$ and $\beta$. Based on this, it can be concluded that all $\bar{a}$ corresponding to $P^{out}_{rad}(\bar{a}) = \bar{a}^H \cdot \bar{\bar{P}}_+ \cdot \bar{a} = 0$ constitute a linear space [16], and this space is called as non-radiative space in this paper.

Any non-radiative mode has a zero radiation pattern at infinity due to the second equality in (2.1), so its fields are zeros at any point out of scatterer based on the conclusions given in [22], and then its tangential electric field is zero on whole $\partial V$. This implies that $\bar{\bar{Z}} \cdot \bar{a} = 0$ for any non-radiative mode, here the $\bar{\bar{Z}}$ is the impedance matrix derived from the Galerkin-based EFIE-MoM. The interior resonance space constructed by all interior resonant modes is identical to $\mathbb{N}(\bar{\bar{Z}})$ [23], here the $\mathbb{N}(\bar{\bar{Z}})$ is the zero space of $\bar{\bar{Z}}$, and then any non-radiative mode must be the interior resonant mode. In addition, it is obvious that any interior resonant mode is non-radiative.

Based on above these observations and considering of that $\bar{\bar{P}} = (1/2)\bar{\bar{Z}}$, it can be concluded that the non-radiative space, the interior resonance space, the zero space $\mathbb{N}(\bar{\bar{P}}_+)$, the zero space $\mathbb{N}(\bar{\bar{P}})$, and the zero space $\mathbb{N}(\bar{\bar{Z}})$ are identical to each other, and all the CMs corresponding to $P^{out}_{rad}(\bar{a}) = 0$ constitute a orthogonal basis of these spaces.

Based on the above these, the normal EMT for closed PEC structures is classified into the PEC-EMP-CMT framework.

### B. The discussions for various electromagnetic powers and various CMs.

The total stored energy $\{W^{sca}_e, W^{sca}_m\}$ includes two parts, the non-radiative/reactive stored energy $\{W^{sca,react}_e, W^{sca,react}_m\}$ and the radiative stored energy $\{W^{sca,rad}_e, W^{sca,rad}_m\}$, and $W^{sca,rad}_e = W^{sca,rad}_m$ [15], so

$$P^{inp}_{react} = P^{out}_{sto} = 2\omega(W^{sca}_m - W^{sca}_e)$$
$$= 2\omega(W^{sca,react}_m - W^{sca,react}_e) \quad (68)$$

This is just the common reason why the $P^{out}_{sto}$ is called as "reactively stored power" (simply as "stored power") in this paper, and why the subscript "$sto$" is utilized in the symbol $P^{out}_{sto}$.

For the radiation systems, the most desired operating state is that $W^{sca,react}_m = 0 = W^{sca,react}_e$ [15], [21], so the $P^{out}_{sto}(\bar{J}^{sca}) = 0$ corresponding to non-zero $\bar{J}^{sca}$ is usually desired in antenna engineering community. However, that $P^{out}_{sto}(\bar{J}^{sca}) = 0$ is only a necessary condition to guarantee that $W^{sca,react}_m = 0 = W^{sca,react}_e$ instead of the sufficient condition, and an extreme example is the interior resonant state of PEC cavity. In fact, this is just the reason why sometimes the strongest radiation state and the resonant state cannot be simultaneously archived as explained in following Sec. VIII-C.

Based on above observations, it is found out that to minimize the absolute value of the functional $F_3(\bar{J}^{sca})$ in (43) is neither the necessary condition nor the sufficient condition to guarantee that the system operates at the strongest radiation state, so it is valuable to develop the RaCM set for optimizing system radiated power.

For the wireless energy transfer systems [24], the most suitable one in various energies is the $W^{sca,react}_m$, and the reasons are that: (1) the $W^{sca,react}_m$ is not radiative, so this kind of energy is stored in near field zone and not dissipated, and then it is valuable for wireless energy transfer systems to increase transmission efficiency; (2) the $W^{sca,react}_m$ is magnetic instead of being electric, so this kind of energy is more safe for creatures. In fact, the system's strongest state to store reactive magnetic energy can be found in the StoCM set as discussed in Sec. V.

Some aspects of the small antenna applications of the OutCM set, which has the same physical essence as the CM set constructed in PEC-SEFIE-CMT [2]-[3], have been comprehensively reviewed in [8], so they are not repeated here. In addition, the OutCM set constructed in this paper can also be applied to the electrically large antennas (such as the travelling wave antennas [25]) and the non-radiative PEC structures (such as the PEC cavities [11]).

### C. The relation between resonance and the strongest radiation.

Based on the formulations in (5), to maximize $P^{out}_{rad}$ and to maximize $P^{inp}_{act}$ are equivalent to each other, i.e., the most strongest radiation state of scattering current $\bar{J}^{sca}$ and the most efficient way that the power is done by incident field $\bar{E}^{inc}$ on scatterer are equivalent to each other, and the latter state as a representative is discussed in this subsection. To efficiently distinguish the time domain quantities and the frequency domain quantities from each other, the time variable $t$ is specifically added to the time domain quantities in this subsection.

For any point $\bar{r} \in \partial V$, the time average of $P^{inp}(\bar{r},t) = \bar{J}^{sca}(\bar{r},t) \cdot \bar{E}^{inc}(\bar{r},t)$ is as follows

$$\frac{1}{T}\int_0^T P^{inp}(\bar{r},t)\,dt = \sum_{i=x,y,z} \text{Re}\left\{\frac{1}{2}\left[J_i^{sca}(\bar{r})\right]^* E_i^{inc}(\bar{r})\right\}$$

$$= \sum_{i=x,y,z} \frac{1}{2} J_{i0}^{sca}(\bar{r}) E_{i0}^{inc}(\bar{r}) \cos\Delta\varphi_i(\bar{r})$$

$$\leq \sum_{i=x,y,z} \frac{1}{2} J_{i0}^{sca}(\bar{r}) E_{i0}^{inc}(\bar{r}) \quad (69)$$

$$= \frac{1}{2} J_0^{sca}(\bar{r}) E_0^{inc}(\bar{r}) \hat{e}^J(\bar{r}) \cdot \hat{e}^E(\bar{r})$$

$$\leq \frac{1}{2} J_0^{sca}(\bar{r}) E_0^{inc}(\bar{r})$$

here $T=1/f$; the $E_{i0}^{inc}$ and $J_{i0}^{sca}$ are the magnitudes of $E_i^{inc}$ and $J_i^{sca}$; $\Delta\varphi_i = \varphi_i^E - \varphi_i^J$, and the $\varphi_i^E$ and $\varphi_i^J$ are respectively the phases of $E_i^{inc}$ and $J_i^{sca}$; the $\hat{e}^E$ and $\hat{e}^J$ are the direction vectors of $\bar{E}_0^{inc}$ and $\bar{J}_0^{sca}$; $\bar{E}_0^{inc} = \sum_{i=x,y,z} \hat{i} E_{i0}^{inc}$, and $E_0^{inc} = (\bar{E}_0^{inc} \cdot \bar{E}_0^{inc})^{1/2}$; the $\bar{J}_0^{sca}$ and $J_0^{sca}$ can be similarly expressed.

Based on the (69) and that $J_0^{sca} E_0^{inc} \geq 0$, the following relation (70) is derived.

$$P_{rad}^{out} = P_{act}^{inp} = \text{Re}\left\{P^{inp}\right\} = \frac{1}{T}\int_0^T \left[\iint_{\partial V} P^{inp}(\bar{r},t)\,dS\right]dt$$

$$= \iint_{\partial V}\left[\frac{1}{T}\int_0^T P^{inp}(\bar{r},t)\,dt\right]dS \quad (70)$$

$$\leq \frac{1}{2}\iint_{\partial V} J_0^{sca}(\bar{r}) E_0^{inc}(\bar{r})\,dS$$

In (69), the first inequality takes the mark of equality, if and only if $\Delta\varphi_i(\bar{r})=0$ for any $i=x,y,z$, i.e., the $E_i^{inc}(\bar{r},t)$ and $J_i^{sca}(\bar{r},t)$ have the same initial phase for any $i=x,y,z$; the second inequality takes the mark of equality, if and only if $\hat{e}^J(\bar{r})\cdot\hat{e}^E(\bar{r})=1$, i.e., $\bar{E}_0^{inc}(\bar{r})\|\bar{J}_0^{sca}(\bar{r})$. Obviously, above two conditions are equivalent to that $\bar{E}^{inc}(\bar{r},t)\|\bar{J}^{sca}(\bar{r},t)$ at any time $t$ for a certain point $\bar{r}$. In (70), the inequality takes the mark of equality, if and only if $\bar{E}^{inc}(\bar{r},t)\|\bar{J}^{sca}(\bar{r},t)$ holds for any $(\bar{r},t)$, and this corresponds to the most efficient way that the $\bar{E}^{inc}$ works on $\bar{J}^{sca}$, i.e., the most strongest radiation state of $\bar{J}^{sca}$.

In addition,

$$P_{sto}^{out} = P_{react}^{inp} = \text{Im}\left\{P^{inp}\right\} = \text{Im}\left\{\iint_{\partial V}\frac{1}{2}\left[\bar{J}^{sca}(\bar{r})\right]^* \cdot \bar{E}^{inc}(\bar{r})\,dS\right\}$$

$$= \iint_{\partial V} \text{Im}\left\{\frac{1}{2}\left[\bar{J}^{sca}(\bar{r})\right]^* \cdot \bar{E}^{inc}(\bar{r})\right\}dS \quad (71)$$

$$= \iint_{\partial V}\left[\sum_{i=x,y,z}\frac{1}{2} J_{i0}^{sca}(\bar{r}) E_{i0}^{inc}(\bar{r}) \sin\Delta\varphi_i(\bar{r})\right]dS$$

Based on the (69)-(71), it is clear that the most efficient way that the scatterer radiates needs that $\sin\Delta\varphi_i(\bar{r})=0$ for any $\bar{r}\in\partial V$ and any $i=x,y,z$, and then needs that $P_{sto}^{out}=P_{react}^{inp}=\text{Im}\{P^{inp}\}=0$ (i.e., resonance). However, the resonance is not the sufficient condition to guarantee the most efficient radiation state, because it cannot be guaranteed that the $\sin\Delta\varphi_i(\bar{r})$ has the same sign for any $\bar{r}\in\partial V$ and any $i=x,y,z$.

From above observations, the following conclusions are derived.

1) The most complete descriptions for the scatterer's ability to radiate energies (i.e., the effectiveness of the interaction between $\bar{E}^{inc}$ and scatterer) are the distribution of $\cos\Delta\varphi_i(\bar{r})$ and the distribution of $(1/2)J_{i0}^{sca}(\bar{r})E_{i0}^{inc}(\bar{r})$ on $\partial V$, i.e., the distribution of $\text{Re}\{(1/2)[J_i^{sca}(\bar{r})]^* E_i^{inc}(\bar{r})\}$ on $\partial V$, here $i=x,y,z$.

Of cause, the above mathematical descriptions can also be more physically depicted as follows. The effectiveness of the interaction between $\bar{E}^{inc}$ and $\bar{J}^{sca}$ depends on that: (1.1) the distributions of the phase relations between the components of $\bar{E}^{inc}$ and $\bar{J}^{sca}$; (1.2) the distributions of the magnitudes of $\bar{E}^{inc}$ and $\bar{J}^{sca}$; (1.3) the distributions of the polarizations of $\bar{E}^{inc}$ and $\bar{J}^{sca}$.

2) The resonance condition is only a necessary condition to guarantee the strongest radiation of electromagnetic system, but not a sufficient condition.

*D. The discussions and comparisons for the PEC-SEFIE-CMT and PEC-EMP-CMT.*

In this subsection, the core principle of PEC-SEFIE-CMT is summarized at first, and then some necessary discussions for PEC-SEFIE-CMT are provided.

**The core principle of PEC-SEFIE-CMT**

The PEC-SEFIE-CMT was built based on the EFIE-based MoM in [2]-[3].

1) The main destination of PEC-SEFIE-CMT is to construct a transformation from any mathematically complete basis function set $\{\bar{b}_\xi\}_{\xi=1}^\Xi$ to the CM set $\{\bar{a}_\xi\}_{\xi=1}^\Xi$ which satisfies the power orthogonality given by the formulations (17) and (22)-(24) in paper [2].

2) The PEC-SEFIE-CMT achieves the above destination by decomposing the impedance matrix $\bar{\bar{Z}}$ in terms of its real part $\bar{\bar{R}}$ and imaginary part $\bar{\bar{X}}$, and then solving the generalized characteristic equation $\bar{\bar{X}}\cdot\bar{a}=\lambda\bar{\bar{R}}\cdot\bar{a}$.

3) In fact, realizing the destination in 1) by using the method in 2) relies on the following necessary conditions.

(3.1) $P^{out}(\bar{a})=\bar{a}^H\cdot\bar{\bar{Z}}\cdot\bar{a}$;

(3.2) The $\bar{\bar{R}}$ and $\bar{\bar{X}}$ must be symmetric matrices to guarantee that $\text{Re}\{P^{out}(\bar{a})\}=\bar{a}^H\cdot\bar{\bar{R}}\cdot\bar{a}$ and $\text{Im}\{P^{out}(\bar{a})\}=\bar{a}^H\cdot\bar{\bar{X}}\cdot\bar{a}$;

(3.3) The $\bar{\bar{R}}$ must be positive definite [16].

4) To guarantee the conditions listed in 3), the PEC-SEFIE-CMT is realized as follows.

(4.1) To guarantee the condition (3.1), the $\bar{\bar{Z}}$ must be constructed by using inner product, and the basis function set and the testing function set must be the same, and a coefficient "1/2" should be contained;

(4.2) To guarantee the condition (3.2), the $\bar{\bar{Z}}$ must be constructed by using symmetric product, and the basis function set and the testing function set must be the same;

(4.3) To guarantee the condition (3.3), the scatterer is required to radiate some electromagnetic energy.

To simultaneously achieve above (4.1) and (4.2), the PEC-SEFIE-CMT expands the scattering current in terms of real basis function set [3], and then the PEC-SEFIE-CMT can only construct the real characteristic currents, because in the vector space $\mathbb{R}^\Xi$ the characteristic vectors derived from $\bar{\bar{X}}\cdot\bar{a}=\lambda\bar{\bar{R}}\cdot\bar{a}$ are real.

**The discussions for PEC-SEFIE-CMT**

Firstly, the PEC-SEFIE-CMT has not ability to provide the complex characteristic currents, but it is more suitable for some electrically large structures, such as the travelling wave antennas, to depict their inherent characteristics by using the complex characteristic currents [25].

Secondly, the PEC-SEFIE-CMT doesn't provide any efficient method to research the non-radiative modes. Although some scholars have found out that the CMs corresponding to the infinite characteristic values are just the non-radiative modes [2], [26]-[27], the infinite characteristic values are not the necessary conditions to judge the non-radiative modes as explained in the Sec. VI of this paper.

**The discussions for the OutCM set in PEC-EMP-CMT**

At first, it must be clearly pointed out here that the physical essence of the OutCM set constructed in this paper is the same as the CM set derived from the traditional PEC-SEFIE-CMT. However, the former is more advantageous than the latter in the following aspects.

In this paper, the output power matrix $\bar{\bar{P}}$ is decomposed in terms of two Hermitian matrices, $\bar{\bar{P}}_+$ and $\bar{\bar{P}}_-$, as illustrated in the formulation (10), and they respectively correspond to the radiated power and stored power as proven in the (11). This decomposition can always be realized, and doesn't need to restrict the basis function set $\{\bar{b}_\xi\}_{\xi=1}^\Xi$ to be real. Based on this, the characteristic currents are not restricted to be real, and then this paper provides a possible way for constructing the complex characteristic currents.

When the $\bar{\bar{P}}_+$ is positive definite, the OutCM set is obtained by solving the generalized characteristic equation $\bar{\bar{P}}_- \cdot \bar{a} = \lambda \bar{\bar{P}}_+ \cdot \bar{a}$ in (40), and this equation is the necessary condition to orthogonalize the modal output powers as (45). When the $\bar{\bar{P}}_+$ is positive semi-definite at frequency $f_0$, the frequency $f_0$ is efficiently recognized by employing the modal output impedance $Z_\xi^{out}$ given in (41), and then the OutCM set at $f_0$ is obtained by using a "limiting method" given in (44).

In addition, the PEC-EMP-CMT is not only suitable for the scatterer placed in vacuum like PEC-SEFIE-CMT, but also valid for the scatterer placed in any time-harmonic electromagnetic environment. It is found out in this paper that for a scatterer, which is regarded as a whole object, its various power-based CM sets are independent of the electromagnetic environment which surrounds it.

**The relation between PEC-SEFIE-CMT and PEC-EMP-CMT**

Based on the above observations, the CM set derived from PEC-SEFIE-CMT can be viewed as a special case of the OutCM set in PEC-EMP-CMT framework, when the following conditions are simultaneously satisfied. (a) Only real characteristic currents are desired; (b) there doesn't exist any non-radiative mode; (c) the scatterer is placed in vacuum.

In fact, a more valuable effect to treat the PEC-SEFIE-CMT as a special case of PEC-EMP-CMT is to liberate the PEC-SEFIE-CMT from its traditional framework, IE-MoM, and this liberation is very important for CMT, as discussed in the following Sec. VIII-E.

*E. The discussions for the variants of PEC-SEFIE-CMT under the IE-MoM framework.*

After that the physical validity of the PEC-SEFIE-CMT is verified by some applications in antenna community [28]-[30], some IE-MoM-based variants of PEC-SEFIE-CMT are raised gradually.

Obviously, only the impedance matrix derived from the Galerkin-based EFIE-MoM for PEC system has the same physical essence as the output power quadratic matrix as discussed in Sec. VIII-D. It has been pointed out in [4] and [20] that the power characteristic of the CM set derived from Mat-VIE-CMT is physically unreasonable.

In fact, the physical validities of all variants for the PEC-SEFIE-CMT under IE-MoM framework should be reexamined carefully.

*F. The discussions for the modal quantities introduced in this paper and the traditional characteristic quantity Modal Significance (MS).*

Obviously, the various modal quantities discussed in the Sec. VII and the modal component $|\tilde{c}_\xi^{out}|^2 \tilde{P}_\xi^{out}$ in the expansion formulation (60) satisfy the relation (72).

In fact, the $\text{MAOP}_\xi^{mod,\,act}$ in (66) is equivalent to the traditional MS as illustrated in the following (73).

$$\begin{aligned}
\text{Traditional MS}_\xi &\triangleq \frac{1}{\left|1 + j\,\lambda_\xi^{out}\right|} \\
&= \frac{1}{\left|1 + j\left(\text{Im}\{\tilde{P}_\xi^{out}\}\big/\text{Re}\{\tilde{P}_\xi^{out}\}\right)\right|} \\
&= \frac{\text{Re}\{\tilde{P}_\xi^{out}\}}{\left|\text{Re}\{\tilde{P}_\xi^{out}\} + j\,\text{Im}\{\tilde{P}_\xi^{out}\}\right|} \\
&= \frac{\text{Re}\{\tilde{P}_\xi^{out}\}}{\left|\tilde{P}_\xi^{out}\right|} \\
&= \text{MAOP}_\xi^{mod,\,act}
\end{aligned} \quad (73)$$

here the second equality is due to the relation (42), and the third equality originates from that $\text{Re}\{P_\xi^{out}\} > 0$ for the radiative OutCMs.

It is easily found out from (72)-(73) that the physical essence of the traditional MS is to quantitively depict the modal ability to allocate the total modal output power to its active part, instead of a quantitive depiction for the modal weight in whole modal expansion formulation

The essential reason leading to the above problem of the traditional MS is carefully analyzed as below.

When the CM $M$ is normalized by using the normalization way given in [2], i.e., the modal radiated power is normalized to be unit, the normalized CM is specifically denoted as the symbol $\breve{M}$ to be distinguished from the normalized version $\tilde{M}$

$$\left|\tilde{c}_\xi^{out}\right|^2 \tilde{P}_\xi^{out} = \text{SMS}_\xi^{sys,\,act} + j\,\text{SMS}_\xi^{sys,\,react} = \overbrace{\text{MACE}_\xi^{mod} \cdot \text{GMS}_\xi^{sys,\,tot}}^{\text{SMS}_\xi^{sys,\,tot}} \cdot \text{MAOP}_\xi^{mod,\,act} + j\,\overbrace{\text{MACE}_\xi^{mod} \cdot \text{GMS}_\xi^{sys,\,tot}}^{\text{SMS}_\xi^{sys,\,tot}} \cdot \text{MAOP}_\xi^{mod,\,react} \\
= \text{MACE}_\xi^{mod} \cdot \underbrace{\text{GMS}_\xi^{sys,\,tot} \cdot \text{MAOP}_\xi^{mod,\,act}}_{\text{GMS}_\xi^{sys,\,act}} + j\,\text{MACE}_\xi^{mod} \cdot \underbrace{\text{GMS}_\xi^{sys,\,tot} \cdot \text{MAOP}_\xi^{mod,\,react}}_{\text{GMS}_\xi^{sys,\,react}} \quad (72)$$

used in this paper. In addition, the $\bar{M}$ version of (60) is the following (60').

$$P^{inp} = P^{out} = \sum_{\xi=1}^{\Xi} \left|\tilde{c}_\xi^{out}\right|^2 \tilde{P}_\xi^{out}$$
$$= \sum_{\xi=1}^{\Xi} \frac{1}{\left|\tilde{P}_\xi^{out}\right|} \cdot \left|\frac{1}{2}\left\langle \tilde{\bar{J}}_\xi^{out}, \bar{E}^{inc} \right\rangle_{\partial V}\right|^2 \cdot \frac{\tilde{P}_\xi^{out}}{\left|\tilde{P}_\xi^{out}\right|} \quad (60')$$

here [2]

$$\text{Re}\left\{\tilde{P}_\xi^{out}\right\} = 1 \quad (74.1)$$
$$\text{Im}\left\{\tilde{P}_\xi^{out}\right\} = \lambda_\xi^{out} \quad (74.2)$$

and

$$\text{Traditional MS}_\xi = \frac{1}{\left|\tilde{P}_\xi^{out}\right|} \quad (75)$$

Obviously, the magnitude of the $\tilde{P}_\xi^{out}/|\tilde{P}_\xi^{out}|$ in (60') is unit just like the term $\tilde{P}_\xi^{out}/|\tilde{P}_\xi^{out}|$ in (60).

However the $\tilde{\bar{J}}_\xi^{out}$ in term $\left|(1/2)\left\langle \tilde{\bar{J}}_\xi^{out}, \bar{E}^{inc}\right\rangle_{\partial V}\right|^2$ is not well normalized. For example, the following case may be existed. $(1/2)\left\langle \tilde{\bar{J}}_\xi^{out}, \tilde{\bar{J}}_\xi^{out}\right\rangle_{\partial V} \gg (1/2)\left\langle \tilde{\bar{J}}_\zeta^{out}, \tilde{\bar{J}}_\zeta^{out}\right\rangle_{\partial V}$, though $\text{Re}\{\tilde{P}_\xi^{out}\} = 1 = \text{Re}\{\tilde{P}_\zeta^{out}\}$. In fact, this problem will not exist for the normalization way used in this paper, because

$$\frac{1}{2}\left\langle \tilde{\bar{J}}_\xi^{out}, \tilde{\bar{J}}_\xi^{out}\right\rangle_{\partial V} = 1 = \frac{1}{2}\left\langle \tilde{\bar{J}}_\zeta^{out}, \tilde{\bar{J}}_\zeta^{out}\right\rangle_{\partial V} \quad (76)$$

for any $\xi, \zeta = 1, 2, \cdots, \Xi$.

In fact, the most essential reason to lead to the above problem in the traditional MS is that the normalization way used in [2] only focuses on the modal radiated powers, but ignores the modal currents and modal fields. However, the normalization way used in (56) focuses on the modal currents, and then the modal fields and modal powers are automatically normalized as the (57) and (58), because the modal fields and modal powers are generated by the modal currents.

*G. Discussions and comparisons for the different ways to normalize electromagnetic quantities.*

The essential characteristics of an electromagnetic quantity will not suffer any influence, when the quantity is multiplied by a non-zero constant scalar, especially for the modal problems [1]-[12]. This is the main reason why a normalized quantity is more desired when some quantitative comparisons are performed, such as the normalized CMs.

In general, there exist many different ways to normalize various quantities. For example, the modal radiated powers are normalized to be unit in PEC-SEFIE-CMT [2]; in this paper, another power normalization way is provided in (6). Some discussions for various normalization ways have been given in the above Sec. VIII-F, and some further discussions are provided in this subsection.

1) The destination to normalize various electromagnetic powers by utilizing the way given in (6).

To normalize various powers as (6) is to efficiently remove the influence from the amplitude of $\bar{J}^{sca}$ and to take account of the distributions of $\bar{J}^{sca}$ and then to efficiently construct the RaCM set, the StoCM set, and the non-radiative OutCMs.

2) The formulation (6) and (56) are consistent with each other.

In this paper, the normalized versions of current and power are separately defined in (56) and (6). In fact, the current and power are related quantities, because the current is the source to generate power. Based on these observations, when the current is artificially normalized, the normalized power will be automatically determined, so the normalization way for power should not be redefined. However, the normalized power is defined in Sec. II, and the normalized current isn't introduced until the Sec. VII-B. The reason is that this paper wants to emphasize that the introductions for the normalized power and the normalized current are based on different destinations as discussed in the Sec. VIII-F and the Sec. VIII-G 1).

Although the normalized power and current are separately defined in this paper, it is obvious that the two definitions are self-consistent, i.e., the power generated by normalized current defined in (56) is just the normalized power defined in (6).

Based on above these, it can be concluded that the normalization way in (6) is equivalent to (56).

3) The physical reasonableness to normalize electromagnetic powers by utilizing the way given in (6).

Some aspects of the physical reasonableness of (6) have been discussed in Sec. VIII-F.

In addition, by analyzing the dimensions of the normalized powers given in this paper, it is easy to find out that the dimensions of $\tilde{P}^{out}$, $\tilde{P}_{rad}^{out}$, and $\tilde{P}_{sto}^{out}$ are Ohm, and then the physical essences of $\tilde{P}^{out}$, $\tilde{P}_{rad}^{out}$, and $\tilde{P}_{sto}^{out}$ are respectively the same as the impedance, resistance, and reactance. Obviously, this is consistent with the tradition of stationary circuit theory, because the concept "resistance" in stationary circuit theory is just the power normalized by the square of current.

Considering of the formulation (6), the $\tilde{P}^{inp} = \tilde{P}^{out}$, $\tilde{P}_{act}^{inp} = \tilde{P}_{rad}^{out}$, and $\tilde{P}_{react}^{inp} = \tilde{P}_{sto}^{out}$ can be respectively defined as the input impedance, input resistance, and input reactance of PEC system, and respectively denoted as follows

$$Z^{inp}(\bar{J}^{sca}) = \begin{cases} \dfrac{P^{inp}(\bar{J}^{sca})}{(1/2)\langle \bar{J}^{sca}, \bar{J}^{sca}\rangle_{\partial V}} &, (\bar{J}^{sca} \neq 0) \\ 0 &, (\bar{J}^{sca} = 0) \end{cases} \quad (77.1)$$

$$R^{inp}(\bar{J}^{sca}) = \begin{cases} \dfrac{P_{act}^{inp}(\bar{J}^{sca})}{(1/2)\langle \bar{J}^{sca}, \bar{J}^{sca}\rangle_{\partial V}} &, (\bar{J}^{sca} \neq 0) \\ 0 &, (\bar{J}^{sca} = 0) \end{cases} \quad (77.2)$$

$$X^{inp}(\bar{J}^{sca}) = \begin{cases} \dfrac{P_{react}^{inp}(\bar{J}^{sca})}{(1/2)\langle \bar{J}^{sca}, \bar{J}^{sca}\rangle_{\partial V}} &, (\bar{J}^{sca} \neq 0) \\ 0 &, (\bar{J}^{sca} = 0) \end{cases} \quad (77.3)$$

here

$$Z^{inp}\left(\bar{J}^{sca}\right) = R^{inp}\left(\bar{J}^{sca}\right) + j\,X^{inp}\left(\bar{J}^{sca}\right) \quad (78)$$

In fact, the new definition for input impedance as (77) is completely based on the language of field instead of basing on the language of circuit (i.e., doesn't need the concept "potential difference"), and completely independent of the external excitation instead of depending on the position of delta-gap source.

4) The normalization way used in (6) is widely applicable.

The normalization way in (6) is not only suitable for both system quantities and modal quantities, but also valid for both radiative and non-radiative modes. However, the normalization way in the traditional PEC-SEFIE-CMT [2] is not applicable to the non-radiative modes, because the zero radiated power cannot be normalized to one.

5) The normalization way used in (6) is generalizable.

When the external excitation source doesn't distribute on the boundary surface of PEC scatterer, the scattering field satisfies the following boundary condition [13], [18].

$$\bar{J}^{sca}(\bar{r}) = \hat{n}^+(\bar{r}) \times \Delta\bar{H}^{sca}(\bar{r}')\big|_{\bar{r}' \to \bar{r}}\,, \quad (\bar{r} \in \partial V) \quad (79.1)$$

$$\frac{\rho^{sca}(\bar{r})}{\varepsilon_0} = \hat{n}^+(\bar{r}) \cdot \Delta\bar{E}^{sca}(\bar{r}')\big|_{\bar{r}' \to \bar{r}}\,, \quad (\bar{r} \in \partial V) \quad (79.2)$$

here $\rho^{sca}$ is the density of surface charge, and

$$\Delta\bar{H}^{sca}(\bar{r}') = \bar{H}^{sca}_+(\bar{r}') - \bar{H}^{sca}_-(\bar{r}')\,, \quad (\bar{r}' \notin \partial V) \quad (80.1)$$

$$\Delta\bar{E}^{sca}(\bar{r}') = \bar{E}^{sca}_+(\bar{r}') - \bar{E}^{sca}_-(\bar{r}')\,, \quad (\bar{r}' \notin \partial V) \quad (80.2)$$

and the $\hat{n}^+(\bar{r})$ in (79) is the normal vector pointing to the domain "+". The subscript "$\bar{r}' \to \bar{r}$" in (79) represents that the $\bar{r}'$ approaches to the $\bar{r}$ on $\partial V$, but the $\bar{r}'$ is not on the $\partial V$.

Based on (79.1) and considering of the continuity of the normal component of $\bar{H}^{sca}$, the normalized output/input power in (6) can also be viewed as the following equivalent version.

$$\tilde{P}^{out/inp} \triangleq \begin{cases} \dfrac{P^{out/inp}}{(1/2)\langle\Delta\bar{H}^{sca}(\bar{r}'),\Delta\bar{H}^{sca}(\bar{r}')\rangle_{\partial V}}\,, & \text{if } \Delta\bar{H}^{sca}(\bar{r}') \text{ is not always } 0 \\ 0\,, & \text{if } \Delta\bar{H}^{sca}(\bar{r}') \text{ is always } 0 \end{cases} \quad (81)$$

here $\bar{r}' \to \bar{r} \in \partial V$.

In addition, considering of that $\bar{H}^{sca}_\pm(\bar{r}')$ and $\bar{E}^{sca}_\pm(\bar{r}')$ are one-to-one correspondence because of the following relation in source-free domain

$$\begin{aligned}\bar{E}^{sca}_\pm(\bar{r}') &= \dfrac{1}{j\omega\varepsilon_0}\nabla'\times\bar{H}^{sca}_\pm(\bar{r}') \\ \bar{H}^{sca}_\pm(\bar{r}') &= \dfrac{1}{-j\omega\mu_0}\nabla'\times\bar{E}^{sca}_\pm(\bar{r})\end{aligned}\,, \quad (\bar{r}' \notin \partial V) \quad (82)$$

and considering of that the destination of normalization is to remove the influence of the amplitude of $\bar{J}^{sca}$ as explained in Secs. II-B and VII, the system output/input power $P^{out/inp}$ can also be normalized as follows

$$\tilde{P}^{out/inp} \triangleq \begin{cases} \dfrac{P^{out/inp}}{(1/2)\langle\Delta\bar{E}^{sca}(\bar{r}'),\Delta\bar{E}^{sca}(\bar{r}')\rangle_{\partial V}}\,, & \text{if } \Delta\bar{E}^{sca}(\bar{r}') \text{ is not always } 0 \\ 0\,, & \text{if } \Delta\bar{E}^{sca}(\bar{r}') \text{ is always } 0 \end{cases} \\ = \begin{cases} \dfrac{P^{out/inp}}{(1/2)\langle\rho^{sca}/\varepsilon_0,\rho^{sca}/\varepsilon_0\rangle_{\partial V}}\,, & (\rho^{sca} \neq 0) \\ 0\,, & (\rho^{sca} = 0) \end{cases} \quad (83) \\ = \begin{cases} \dfrac{P^{out/inp}}{(1/2)\left\langle\dfrac{\nabla_S \cdot \bar{J}^{sca}}{-j\omega\varepsilon_0},\dfrac{\nabla_S \cdot \bar{J}^{sca}}{-j\omega\varepsilon_0}\right\rangle_{\partial V}}\,, & (\nabla_S \cdot \bar{J}^{sca} \neq 0) \\ 0\,, & (\nabla_S \cdot \bar{J}^{sca} = 0) \end{cases}$$

here $\bar{r}' \to \bar{r} \in \partial V$. In (83), the second equality originates from (79.2) and the continuity of the tangential component of $\bar{E}^{sca}$; the third equality originates from the current continuity equation [13], and the symbol "$\nabla_S \cdot \bar{J}^{sca}$" is the 2-dimensional divergence of surface current $\bar{J}^{sca}$. In (83), the inner product for scalars is defined as $\langle f, g\rangle_\Omega \triangleq \int_\Omega f^*g\,d\Omega$.

Because the dimension of $\tilde{P}^{out/inp}$ is Siemens, the $\tilde{P}^{out/inp}$ can also be called as system input admittance, and equivalently denoted as $Y^{inp}$. The real and imaginary parts of $Y^{inp}$ are respectively called as system input/radiated conductance $G^{inp}$ and system input susceptance $B^{inp}$, and

$$Y^{inp}\left(\bar{J}^{sca}\right) = G^{inp}\left(\bar{J}^{sca}\right) + j\,B^{inp}\left(\bar{J}^{sca}\right) \quad (84)$$

and

$$G^{inp}\left(\bar{J}^{sca}\right) = \text{Re}\{\tilde{P}^{inp}\} = \text{Re}\{\tilde{P}^{out}\} = \tilde{P}^{out}_{rad} \quad (85.1)$$

$$B^{inp}\left(\bar{J}^{sca}\right) = \text{Im}\{\tilde{P}^{inp}\} = \text{Im}\{\tilde{P}^{out}\} = \tilde{P}^{out}_{sto} \quad (85.2)$$

Of course, the modal currents, modal fields, and modal powers can also be similarly normalized as (81) and (83).

*H. The discussions and comparisons for CMT and EMT.*

Under the Eigen-Mode Theory (EMT) framework, there doesn't exist any external excitation. However, the external excitation is not necessarily zero for CMT.

For the time-harmonic electromagnetic problems whose solving domains extend to infinity, the Sommerfeld's radiation condition at infinity [13] signifies that the electromagnetic power is continuously exported from the electromagnetic system to infinity, and then a non-zero input power from external excitation to system is necessary to sustain a stationary oscillation of field. In fact, this is just the reason for frequency domain electromagnetic problem why the singular EMT [12] for infinite solving domain usually leads to some complex external resonance frequencies.

Moreover, the normal modes of closed PEC structures [11] and the non-radiative modes are one-to-one correspondence, and the non-radiative CMs constitute the basis of the non-radiative space as proven in Sec. VIII-A. Based on these above, the normal EMT for closed PEC structures is classified into the PEC-EMP-CMT framework.

In addition, the PEC-EMP-CMT is an object-oriented theory, because the selection for the objective power to be optimized by PEC-EMP-CMT is relatively flexible as illustrated in Secs. IV, V, and VI.

## IX. CONCLUSIONS

An electromagnetic-power-based CMT for PEC systems, PEC-EMP-CMT, is built in this paper. A series of power-based CM sets are constructed in the PEC-EMP-CMT framework. The PEC-EMP-CMT can be applied to the PEC systems which are placed in a complex electromagnetic environment, and for a certain PEC object the various power-based CM sets are independent of the external environment.

The RaCM set has ability to construct the modes which optimize the system radiation. The StoCM set has ability to construct the modes which optimize the system reactive power. The OutCM set has the same physical essence as the traditional CM set constructed in PEC-SEFIE-CMT, but the former generalizes the applicable range of the latter.

A new way to normalize various electromagnetic quantities of PEC systems is introduced in this paper. Based on the new normalization way, some traditional concepts, such as input impedance and admittance etc., are redefined based on the language of field instead of the language of circuit; the traditional characteristic quantity MS is generalized, and some new characteristic and non-characteristic quantities are introduced.

In addition, it is proven in this paper that the non-radiative space is identical to the interior resonant space, and the non-radiative CMs constitute a basis of the space. Based on this, the traditional normal EMT for closed PEC structures is classified into the PEC-EMP-CMT framework. A variational formulation for the external scattering problem of PEC structures is provided based on the conservation law of energy.

## APPENDIXES

The proof for the first equality of (16) can be found out in [16]. Inserting the (10.2) into the first equality of (16), the following relation is derived.

$$P_{rad,\xi}^{out}\delta_{\xi\zeta} = \frac{1}{2}\left(\bar{a}_\xi^{rad}\right)^H \cdot \bar{\bar{P}} \cdot \bar{a}_\zeta^{rad} + \frac{1}{2}\left(\bar{a}_\zeta^{rad}\right)^H \cdot \bar{\bar{P}}^H \cdot \bar{a}_\zeta^{rad}$$
$$= \frac{1}{2}\left(\bar{a}_\xi^{rad}\right)^H \cdot \bar{\bar{P}} \cdot \bar{a}_\zeta^{rad} + \frac{1}{2}\left[\left(\bar{a}_\zeta^{rad}\right)^H \cdot \bar{\bar{P}} \cdot \bar{a}_\xi^{rad}\right]^H \quad \text{(A-1)}$$
$$= \frac{1}{2}\left(\bar{a}_\xi^{rad}\right)^H \cdot \bar{\bar{P}} \cdot \bar{a}_\zeta^{rad} + \frac{1}{2}\left[\left(\bar{a}_\zeta^{rad}\right)^H \cdot \bar{\bar{P}} \cdot \bar{a}_\xi^{rad}\right]^*$$

Considering of the physical meanings of the elements of the matrix $\bar{\bar{P}}$, the (A-1) can be rewritten as follows

$$\begin{aligned}P_{rad,\xi}^{out}\delta_{\xi\zeta} &= (1/2)\left[-\langle\bar{J}_\xi^{rad},\bar{E}_\zeta^{rad}\rangle_{\partial V} - \langle\bar{J}_\zeta^{rad},\bar{E}_\xi^{rad}\rangle_{\partial V}^*\right]\\
&= (1/2)\Big\{(1/2\eta_0)\langle\bar{E}_\xi^{rad},\bar{E}_\zeta^{rad}\rangle_{S_\infty}\\
&\quad + j2\omega\big[(1/4)\langle\bar{H}_\xi^{rad},\mu_0\bar{H}_\zeta^{rad}\rangle_{\mathbb{R}^3} - (1/4)\langle\varepsilon_0\bar{E}_\xi^{rad},\bar{E}_\zeta^{rad}\rangle_{\mathbb{R}^3}\big]\Big\}\\
&\quad +(1/2)\Big\{(1/2\eta_0)\langle\bar{E}_\zeta^{rad},\bar{E}_\xi^{rad}\rangle_{S_\infty}\\
&\quad + j2\omega\big[(1/4)\langle\bar{H}_\zeta^{rad},\mu_0\bar{H}_\xi^{rad}\rangle_{\mathbb{R}^3} - (1/4)\langle\varepsilon_0\bar{E}_\zeta^{rad},\bar{E}_\xi^{rad}\rangle_{\mathbb{R}^3}\big]\Big\}^*\\
&= (1/2)\Big\{(1/2\eta_0)\langle\bar{E}_\xi^{rad},\bar{E}_\zeta^{rad}\rangle_{S_\infty}\\
&\quad + j2\omega\big[(1/4)\langle\bar{H}_\xi^{rad},\mu_0\bar{H}_\zeta^{rad}\rangle_{\mathbb{R}^3} - (1/4)\langle\varepsilon_0\bar{E}_\xi^{rad},\bar{E}_\zeta^{rad}\rangle_{\mathbb{R}^3}\big]\Big\}\\
&\quad +(1/2)\Big\{(1/2\eta_0)\langle\bar{E}_\xi^{rad},\bar{E}_\zeta^{rad}\rangle_{S_\infty}\\
&\quad - j2\omega\big[(1/4)\langle\bar{H}_\xi^{rad},\mu_0\bar{H}_\zeta^{rad}\rangle_{\mathbb{R}^3} - (1/4)\langle\varepsilon_0\bar{E}_\xi^{rad},\bar{E}_\zeta^{rad}\rangle_{\mathbb{R}^3}\big]\Big\}\\
&= (1/2\eta_0)\langle\bar{E}_\xi^{rad},\bar{E}_\zeta^{rad}\rangle_{S_\infty}\end{aligned} \quad \text{(A-2)}$$

and then the (16) is proven.

## ACKNOWLEDGEMENT

This work is dedicated to Renzun Lian's mother, Ms. Hongxia Zou. This work cannot be finished without her constant understanding, support, and encouragement.

In addition, R. Z. Lian would like to thank J. Pan, Z. P. Nie, Y. B. Wen, X. Huang, S. D. Huang, and Y. C. Jiao.

Pan taught Lian some valuable theories related to the electromagnetics and antenna. Nie taught Lian some valuable theories and methods on the modal problem, electromagnetic near field, and computational electromagnetics.

In the early stage of this work, there were some valuable discussions among Wen, X. Huang, and Lian; Wen and X. Huang respectively did some computer simulations and wrote some computer codes to verify the original ideas of Lian.

In the middle and late stages of this work, S. D. Huang and Jiao respectively wrote some computer codes and did some computer simulations to verify Lian's ideas.